\documentclass[twocolumn,prb,aps, superscriptaddress,floatfix]{revtex4-1}
\pdfoutput=1
\usepackage[caption=false]{subfig}
\usepackage{graphicx}
\usepackage{float}
\usepackage{amsmath}
\usepackage{bm}
\usepackage{xcolor}
\usepackage{siunitx}

\usepackage[colorlinks=true, allcolors=blue]{hyperref}
\usepackage{color}

\begin{document}

\title{ Effects of momentum-dependent quasiparticle renormalization on the gap structure of iron-based superconductors}

\author{Shinibali Bhattacharyya}
\affiliation{Department of Physics, University of Florida, Gainesville, FL 32611, USA}

\author{P. J. Hirschfeld}
\affiliation{Department of Physics, University of Florida, Gainesville, FL 32611, USA}

\author{Thomas A. Maier}
\affiliation{Computational Sciences and Engineering Division and Center for Nanophase Materials Sciences, Oak Ridge National Laboratory, Oak Ridge, TN 37831, USA}

\author{Douglas J. Scalapino}
\affiliation{Department of Physics, University of California, Santa Barbara, CA 93106, USA}

\date{\today}
\begin{abstract}
We discuss the influence of momentum-dependent correlations
on the superconducting gap structure in iron-based superconductors. Within the weak coupling approach including self-energy effects at the one-loop spin-fluctuation level, 
we construct a dimensionless pairing strength functional which includes the effects of quasiparticle renormalization. The stationary solution of this equation determines the gap function at $T_c$. The resulting equations represent the simplest generalization of spin fluctuation pairing theory to include the effects of an anisotropic quasiparticle weight. We obtain good agreement with experimentally observed anisotropic gap structures in LiFeAs, indicating that the inclusion of quasiparticle renormalization effects in the existing weak-coupling theories can account for the observed anomalies in the gap structure of Fe-based superconductors.

\end{abstract}


\maketitle
\section{Introduction}

The role of electronic correlation effects on the properties of iron-based superconductors (FeSC) has been explored  from a variety of different theoretical perspectives [\onlinecite{deMedici_review,Bascones_review,Biermann_review,Yin2011,Li_orb_sel_16,Ye_doping_FeSc_14}]. With the five $d$ orbitals of Fe actively participating in low-energy processes, the possibility of orbital selective physics implies that weak correlations may exist in the states of electrons of one orbital type while strong correlations may occur in others. This may lead to substantial differences in quasiparticle weights, interactions, magnetism, and orbital ordering; in addition, Cooper pairing itself can become orbital-selective [\onlinecite{Ogata_selectivepairing,Si_selectivepairing,Yin2014,Sprau2017}]. This many-body mechanism augments the gap anisotropy that is already present in the conventional spin-fluctuation pairing model,  proposed as the dominant source of Cooper pairing in FeSCs [\onlinecite{Scalapino2012,HirschfeldCRAS,Chubukov_review}]. The usual argument leading to $s\pm$ pairing comes from interband pair scattering between electron and hole pockets facilitated by enhanced nesting [\onlinecite{Mazin2008}]. Since the  coherent  $d-$orbital weight varies around any given Fermi surface sheet, gap anisotropy is present [\onlinecite{Maier_anisotropy_2009}]. To the extent certain $d$ orbitals are more incoherent than others due to correlations, pairing in these channels may be further suppressed, leading generally to enhanced gap anisotropy [\onlinecite{Kreisel2017}]. 

The degree of electronic correlation is known to vary considerably across the various families of Fe-based superconductors. Local-density approximation (LDA) + dynamical mean-field theory (DMFT) calculations have suggested that the 111 are considerably more correlated than, e.g., the well-studied 122 materials and that stronger interactions lead to a shrinkage of the inner hole pockets but the size and shape of the electron pockets [\onlinecite{Yin2011},\onlinecite{Lee_etal_Kotliar2012},\onlinecite{Ferber2012}] is maintained. Indeed, LiFeAs is one of several FeSCs known to have a Fermi surface which is quite different from what is predicted from Density functional theory (DFT). Angle-resolved photoemission spectroscopy (ARPES) measurements [\onlinecite{BorisenkoLiFeAs}, \onlinecite{Borisenko12}] show that the $\Gamma$ centered $d_{xz}/d_{yz}$ hole pockets are considerably smaller than DFT predictions, and inclusion of local correlations within a DMFT scheme does not seem to solve this discrepancy. Recently, the Fermi surface was shown to be significantly renormalized via non-local correlations within a Two-particle self-consistent approach [\onlinecite{ZantoutPRL2019}], resembling ARPES findings. 

Previous theoretical attempts [\onlinecite{ Thomale2011,Wang13,Saito14,Ahn14,Yin2014}] to understand the ARPES-determined gap structure [\onlinecite{BorisenkoLiFeAs,Borisenko12,Allan12,Umezawa12}] were based on a phenomenological tight-binding band structure consistent with ARPES data [\onlinecite{Wang13}], i.e., describing the correct spectral positions of the bands. Despite some success in explaining some features  of the gap structure, not all  were  reproduced  properly; in particular, the large gaps on the inner hole pockets were not recovered in the conventional spin fluctuation treatments.  Subsequently,  Ref.[\onlinecite{Saito14}] claimed a good overall fit to experiment by including vertex corrections.

In Ref.[\onlinecite{Kreisel2017}], a phenomenological approach assuming orbital selective renormalization of quasiparticle weights, primarily in the $d_{xy}$ channel,  was used to test the proposition that the discrepancies in calculated gap structures arose from self-energy effects neglected in conventional spin fluctuation theory. The results, applied to LiFeAs, FeSe crystals, and FeSe monolayers showed good agreement with measured magnitudes and anisotropies of gaps from ARPES  [\onlinecite{Borisenko12}] and Bogoliubov Quasiparticle Interference (BQPI) experiments [\onlinecite{Allan12}], and yielded quasiparticle weights  qualititatively consistent with LDA+DMFT and other normal state studies. While this approach lacked a microscopic framework within which to calculate the quasiparticle weights, it showed that quasi-particle renormalization effects were important.

These effects have previously been treated within the Fluctuation Exchange Approximation (FLEX) [\onlinecite{Ikeda2008,AritaIkeda2010,Yanagi2010}] but here we explore a multi-orbital random-phase approximation (RPA) which is more simply related to the results of the phenomenological treatment [\onlinecite{Kreisel2017}]. The approach we discuss in here can be considered as a Fermi surface restricted one-loop FLEX calculation. As such, it does not involve any ad-hoc restrictions and is numerically much less expensive. In our approach, the spin and charge fluctuations are treated within the RPA and their effect on both the paring and the single particle self-energy channels are approximated by  the zero frequency RPA interactions on the Fermi surface.

In Sec.\ref{model}, we introduce the five-orbital tight binding model that we will study along with the multi-orbital RPA effective interactions. Following this, the eigenvalue equation for the pairfield strength and the gap function along with the equation for the single particle renormalization factor are discussed. In Sec.\ref{Results}, we demonstrate the results of solving these equations for LiFeAs, and discuss this in comparison to experiments [Sec.\ref{gapresults}]. In Sec.\ref{gapanalysis} we analyze our results in terms of the quasi-particle renormalization weights and the pairing vertex, and in the following Sec.\ref{spm_symmetry} we discuss the tendency of iron-based systems towards  $s^{\pm}$ pairing symmetry due to electronic correlations. Finally, we note the improvement obtained in predicting the pairing structure when the single particle self-energy is included. In Appendix \ref{appendics}, we give the derivation of the dimensionless pairing strength functional and the resulting stationary eigenvalue equation for the pairing strength and gap function.

\section{Model} \label{model}

We start with a five-orbital tight binding Hamiltonian $H_0$ and include local interactions via Hubbard-Hund terms:
    \begin{align} \label{Hamiltonian}
    \begin{split} 
     H   &= H_0 + H_I \\
        &= \sum_{ij\sigma} \sum_{qt} ( t_{ij}^{tq} - \mu_0\delta_{ij}\delta_{tq} ) c^\dagger_{it\sigma} c_{jq\sigma} \\
        & + U \sum_{it} n_{it\uparrow} n_{it\downarrow} + U' \sum_{i,t<q} \sum_{\sigma \sigma'} n_{it\sigma} n_{iq\sigma'} \\
        & + J \sum_{i,t<q} \sum_{\sigma \sigma'} c^\dagger_{it\sigma} c^\dagger_{iq\sigma'} c_{it\sigma'} c_{iq\sigma}  \\
        & + J' \sum_{i,t\neq q} c^\dagger_{it\uparrow} c^\dagger_{iq\downarrow} c_{it\downarrow} c_{iq\uparrow}
    \end{split}					
    \end{align}
where the interaction parameters $U,U',J,J'$ are given in
the notation of Kuroki \textit{et al.} [\onlinecite{Kuroki2008}] We consider
cases which obey spin-rotation invariance (SRI) through the relations $U'=U-2J$ and $J=J'$. The kinetic energy $H_0$ includes the chemical potential $\mu_0$ and is described by a tight-binding model spanned by five Fe $d$ orbitals $ [ d_{x y},d_{x^2 - y^2}, d_{x z},d_{y z},d_{3z^2 -r^2} ]$. Here $q$ and $t$ are the orbital indices and $i,j$ is the Fe-atom site. The spectral representation of the non-interacting Green's function is:
    \begin{align} \label{Gbare}
    \begin{split} 
    \ G_{t q}^0(\mathbf{k},\omega_{m}) & = \sum_{\mu} \frac{a^{t}_{\mu}(\mathbf{k})a^{q*}_{\mu}(\mathbf{k})}{i\omega_{m} - E_{\mu}(\mathbf{k})} \\
    \end{split}					
    \end{align}
where the matrix elements $a^q_\mu(\mathbf{k}) = \langle q | \mu \mathbf{k} \rangle$ are spectral weights of the Bloch state $| \mu \mathbf{k} \rangle$ with band index $\mu$ and wave vector $\mathbf{k}$ in the orbital basis and $ \omega_{m} = (2m+1)\pi k_B T $ are the fermionic Matsubara frequencies for a given temperature $T$. We will adopt Latin symbols to denote orbital indices and Greek ones to denote band indices, throughout the rest of the discussion.  

The orbitally resolved noninteracting susceptibility is:
    \begin{align} 
    \begin{split} 
    & \chi^0_{pqst}(\mathbf{q},\Omega_m) \\
    &= -\frac{1}{N_k \beta}  \sum_{\mathbf{k}\omega_m} G^0_{tq}(\mathbf{k},\omega_m) G^0_{ps}(\mathbf{k+q},\omega_m+\Omega_m) \\	
    & = -\frac{1}{N_k}  \sum_{\mathbf{k} \mu \nu} \frac{ a^t_\mu(\mathbf{k}) a^{q*}_\mu(\mathbf{k}) a^p_\nu(\mathbf{k+q}) a^{s*}_\nu(\mathbf{k+q})}{i\Omega_m + E_\mu(\mathbf{k}) - E_\nu(\mathbf{k+q})} \\
    & \qquad \qquad \qquad \times [f(E_\mu(\mathbf{k})) - f( E_\nu(\mathbf{k+q})) ]
    \end{split}
    \end{align}
where $ N_k $ is the number of Fe lattice sites and $\beta=1/k_B T$ is the inverse temperature, $ \Omega_m= 2m\pi k_B T $ is the bosonic Matsubara frequency. Within the RPA, the charge-fluctuation and spin-fluctuation parts of the RPA susceptibility are given by:
    \begin{align} 
    \begin{split} \label{RPAsus}
    \chi^C(\mathbf{q},\Omega_m) &=  [ 1 + \chi^0(\mathbf{q},\Omega_m) U^C]^{-1} \chi^0(\mathbf{q},\Omega_m) \\
    \chi^S(\mathbf{q},\Omega_m) &=  [ 1 - \chi^0(\mathbf{q},\Omega_m) U^S]^{-1} \chi^0(\mathbf{q},\Omega_m) 
    \end{split}					
    \end{align}
The interaction matrices $U^C$ and $U^S$ in orbital space have the following elements:
   \begin{align}
   \begin{array}{ll}
	 U^{C}_{pppp}= U     &  ,U^{S}_{pppp}= U  \\
	 U^{C}_{ppss}= 2U'-J &  ,U^{S}_{ppss}= J  \\
	 U^{C}_{pssp}= J'    &  ,U^{S}_{pssp}= J' \\
	 U^{C}_{psps}= 2J-U' &  ,U^{S}_{psps}= U'. \\
   \end{array}
   \end{align}
Defining $ U^{SC} = U^S + U^C$, the particle-hole ($N$) and the singlet pairing particle-particle ($A$) interactions are:
 \begin{align}
 \begin{split} \label{VAint}
     \left[ V_{p q s t}(\mathbf{q},\Omega_{m}) \right]_N &= \left[ \frac{3}{2} U^S \chi^S(\mathbf{q},\Omega_{m}) U^S +  \frac{3U^S}{2}  \right. \\
     & + \frac{1}{2} U^C \chi^C(\mathbf{q},\Omega_{m}) U^C - \left. \frac{U^C}{2}   \right. \\
     & \left. - \frac{1}{4} U^{SC} \chi^0(\mathbf{q},\Omega_{m}) U^{SC} \right]_{pqst} \\
     \left[ V_{p q s t}(\mathbf{q},\Omega_{m}) \right]_A &= \left[ \frac{3}{2} U^S \chi^S(\mathbf{q},\Omega_{m}) U^S +  \frac{U^S}{2}  \right. \\
     & - \frac{1}{2} U^C \chi^C(\mathbf{q},\Omega_{m}) U^C + \left. \frac{U^C}{2}   \right]_{p q s t}.
 \end{split}
 \end{align}
In the following, we assume that the dynamics of the interaction is cut off on a spin-fluctuation energy scale and we will restrict the treatment of the problem to Bloch states on the Fermi surface. We have assumed that the leading pairing instability occurs in the even-frequency channel. Odd-frequency pairing, if it occurs, is expected to be associated with a reduced critical temperature [\onlinecite{Schaffer2020}]. Assuming that no bands cross each other in the vicinity of the Fermi level ( i.e. $\Omega_m = \Omega = 0$ eV), each Fermi momentum point corresponds to a unique band quantum number, i.e. $\mathbf{k} \in \mu$, and $ \mathbf{k'} \in \nu$. The corresponding band representation for the interaction vertices, describing scattering of particles between $(\mathbf{k_\mu,k'_\nu})$ states at the Fermi level is 
   \begin{align}
   \begin{split}
   \left[ V(\mathbf{k_\mu,k'_\nu},0) \right]_N &= \text{Re}  \sum_{pqst} \left( a^{q*}_\mu(\mathbf{k_\mu})a^{s*}_\nu(\mathbf{k'_\nu}) \right. \\
   & \left. \left[ V_{pqst}(\mathbf{k_\mu-k'_\nu},0) \right]_N a^{t}_\mu(\mathbf{k_\mu})a^{p}_\nu(\mathbf{k'_\nu}) \right) \\
   \left[ V(\mathbf{k_\mu,k'_\nu},0) \right]_A &= \text{Re}  \sum_{pqst} \left( a^{p*}_\mu(\mathbf{k_\mu})a^{t*}_\mu(\mathbf{-k_\mu}) \right. \\
   & \left. \left[ V_{pqst}(\mathbf{k_\mu-k'_\nu},0) \right]_A a^{q}_\nu(\mathbf{k'_\nu})a^{s}_\nu(\mathbf{-k'_\nu}) \right) .
   \label{normalvertex}
   \end{split}
   \end{align}
The symmetrized singlet pairing vertex is then given by
\begin{equation} \label{singletvertex}
 \left[ V(\mathbf{k_\mu,k'_\nu},0) \right]_A = \frac{1}{2} \left( \left[ V(\mathbf{k_\mu,k'_\nu},0) \right]_A + \left[ V(\mathbf{-k_\mu,k'_\nu},0) \right]_A\right) .
\end{equation}
As discussed in Appendix \ref{appendics}, the stationary solution of the pairing strength functional (Eq.\ref{lambda_equation}) is determined by the eigenvalue equation: 
\begin{align} 
   \label{Gapeqn}
   \lambda g(\mathbf{k_\mu}) &= -\frac{1}{(2\pi)^d} \oint_{\mathbf{k'_\nu} \in FS } \frac{ \left[ V(\mathbf{k_\mu,k'_\nu},0) \right]_A }{Z(\mathbf{k'_\nu}) } \frac{d^{d-1}k'_\nu}{|\mathbf{v}_F(\mathbf{k'_\nu})|} g(\mathbf{k'_\nu}).
\end{align}
with 
\begin{align} 
   \label{Zeqn}
   Z(\mathbf{k_\mu}) &= 1 + \frac{1}{ (2\pi)^d} \oint_{\mathbf{k'_\nu} \in FS } \left[ V(\mathbf{k_\mu,k'_\nu},0) \right]_N \frac{d^{d-1}k'_\nu}{|\mathbf{v}_F(\mathbf{k'_\nu})|} .
\end{align}
Here, $\mathbf{v}_F(\mathbf{k'_\nu})$ is the Fermi velocity of band $\nu$ and the integration is over its corresponding Fermi surface. The eigenfunction $g(\mathbf{k_\mu}) $ then determines the symmetry and structure of the leading pairing gap close to $T_c$. Traditional spin-fluctuation pairing calculation sets  $Z(\mathbf{k_\mu})$ to unity. Instead, we use the solution for $Z(\mathbf{k_\mu})$ from Eq.\ref{Zeqn} in Eq.\ref{Gapeqn}. Solving Eq.\ref{Gapeqn} for the eigenfunction $g(\mathbf{k_\mu}) $ corresponding to the leading eigenvalue $\lambda$ gives the modified gap function $ \Delta(\mathbf{k_\mu}) = \frac{ g(\mathbf{k_\mu}) }{ Z(\mathbf{k_\mu}) }$. Note that we adopt here the notation $Z({\bf k})$   consistent with Eliashberg's convention, such that $Z\gg 1$ corresponds to a highly incoherent quasiparticle. This is in contrast to the notation adopted in much of the orbital selective literature, where $Z$ rather than $Z^{-1}$ is the quasiparticle weight (see, e.g. Sprau et al. [\onlinecite{Sprau2017}]).

In the results presented here, we perform 2D calculations with a $\mathbf{k}$-mesh on the order of $ 100 \times 100$ in the unfolded Brillouin Zone (BZ), and $\approx 500$ total number of Fermi surface points. We set $k_B T=0.01eV$ for the rest of this paper.

\section{Results} \label{Results}

To study the effects of momentum-dependent correlations on the pairing structure, we have chosen the electronic dispersion relevant to LiFeAs as in Ref.[\onlinecite{Wang13}]. Certain Fermi sheets in LiFeAs are known to have significant $k_z$-dispersion. However, to compare with experimental data of measured gap magnitudes, we only consider the $k_z=\pi$ plane for our analysis, where all the $Z=(0,0,\pi)$-centered hole pockets are present. We stress that the 3D properties of the dispersion do not affect the conclusions of our results, since $\chi({\bf q},0)$ is nearly independent of $q_z$ [\onlinecite{Wang13}].

\begin{figure*}[hbt!]
     \centering
     \subfloat[]{\includegraphics[width=0.297\textwidth]{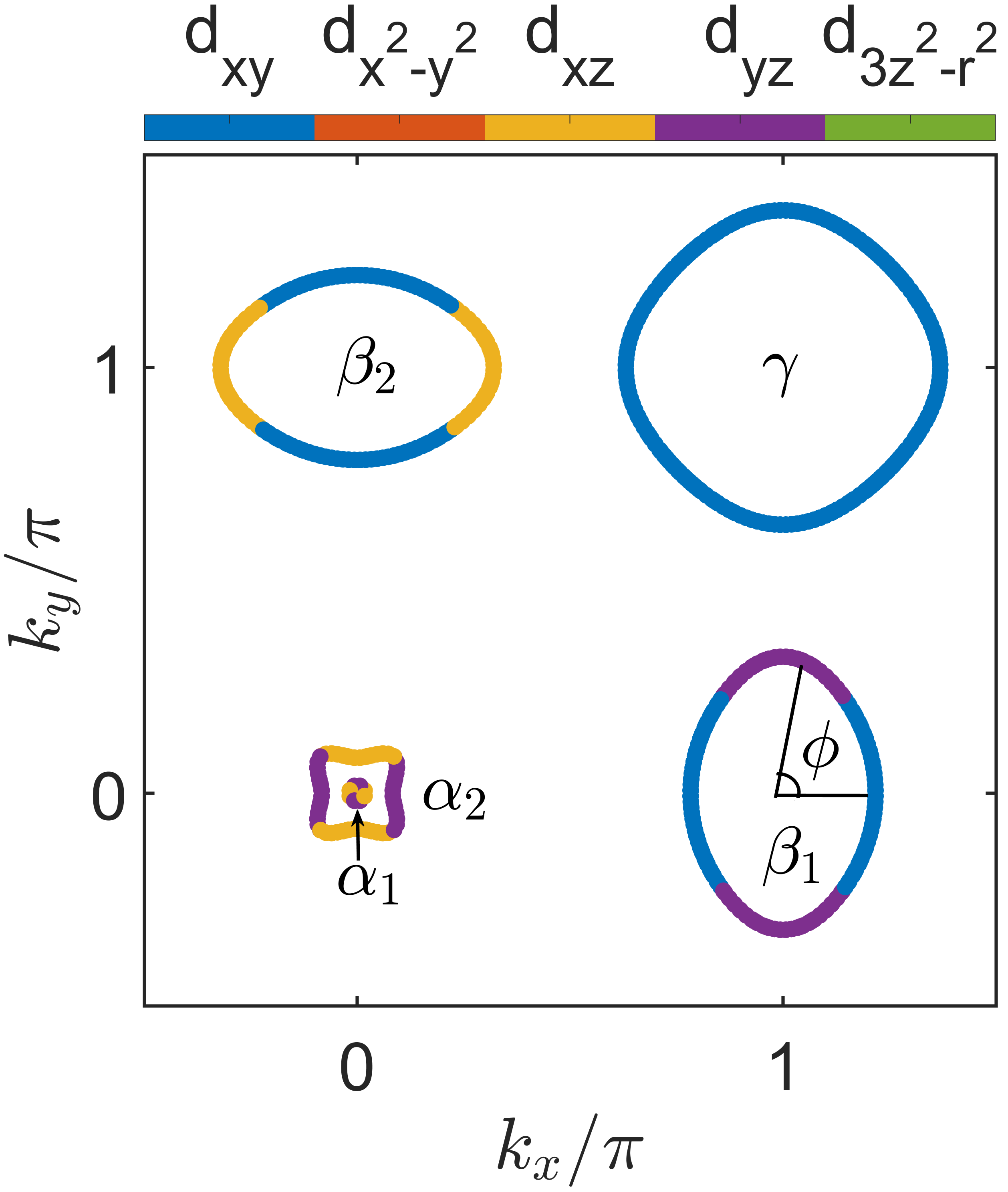}\label{FS}} \,
     \subfloat[]{\includegraphics[width=0.325\textwidth]{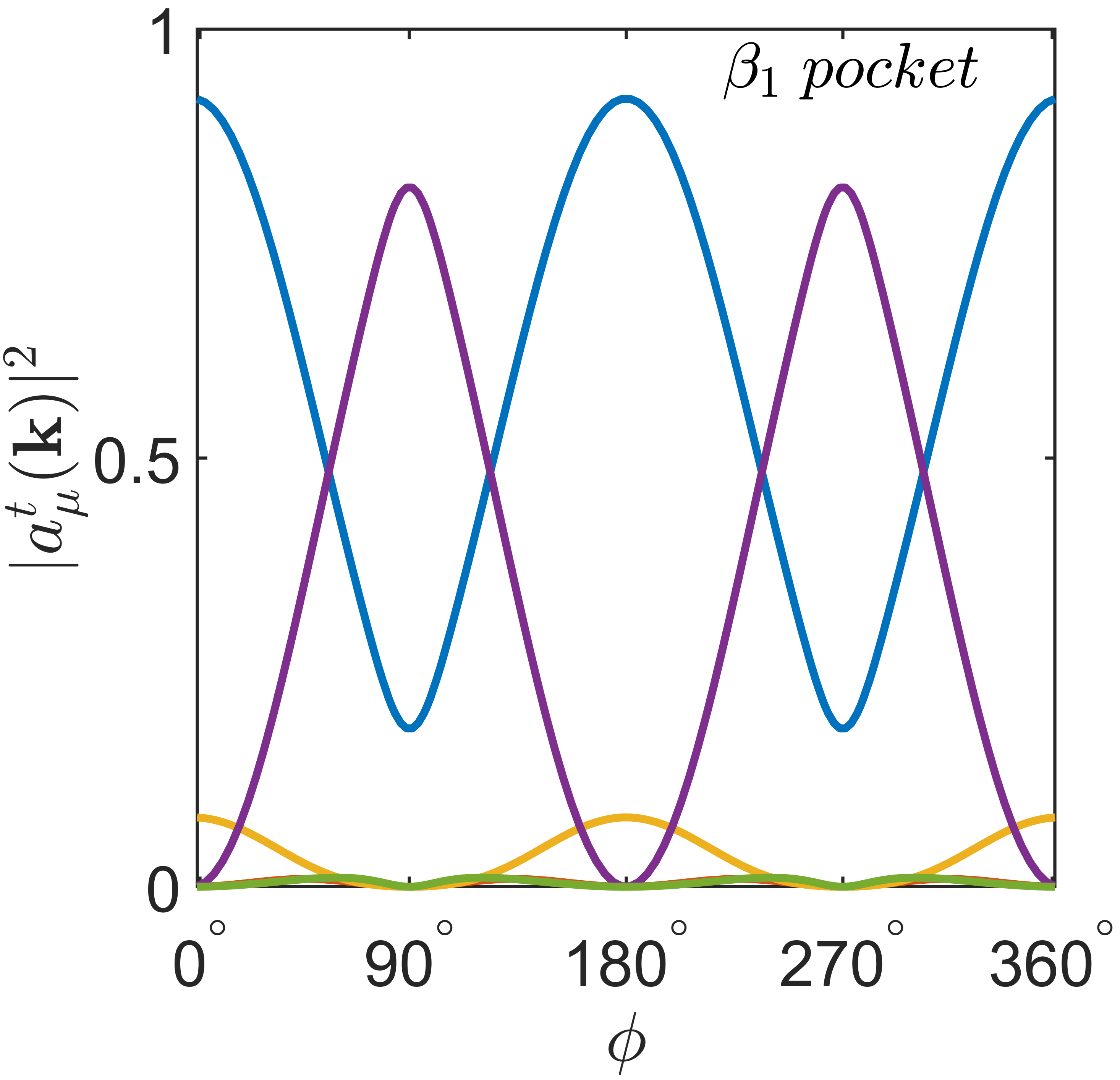}\label{orbwt}}
     \subfloat[]{\includegraphics[width=0.361\textwidth]{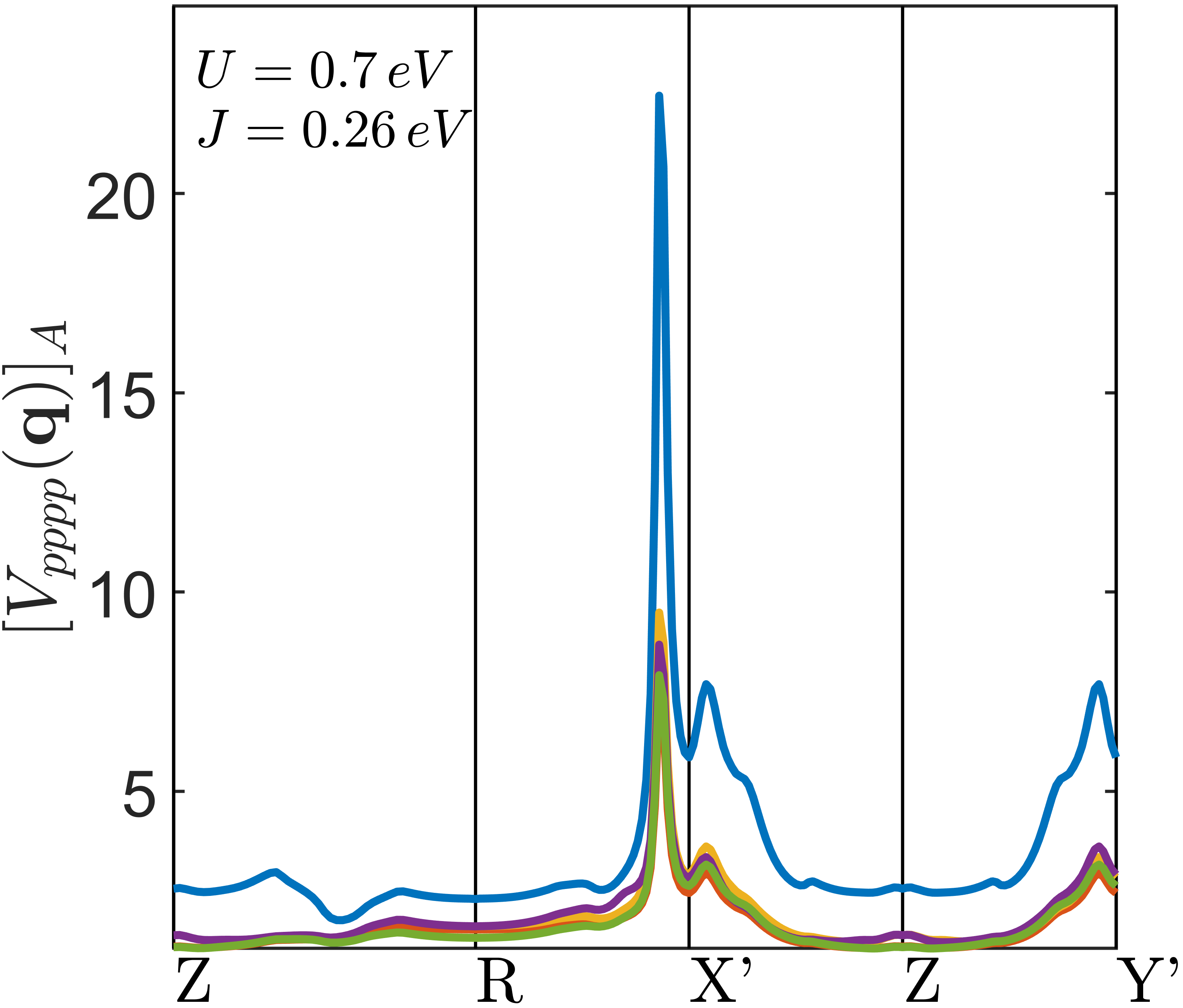}\label{VA_HSP}}
     \caption{(a) Fermi surface at $k_z = \pi$ plane for LiFeAs represented by its corresponding dominant orbital character as indicated in the color legend. (b) Orbital weight along the $\beta_1$ electron pocket plotted as a function of angle $\phi$ shown in (a). (c) Diagonal orbital components of the static particle-particle effective interaction $[V_{pppp}(\mathbf{q},0)]_A$ along high-symmetry path $Z-R-X'-Z-Y'$ with $U$ and $J$ as indicated. Plots as a function of the angle $\phi$ around the Fermi pockets are done with the angle measured counterclockwise from the $k_x$ axis.}
\end{figure*}

As shown in Fig.\ref{FS}, for the undoped material with a filling of $n=6$ in the 1-Fe BZ, the Fermi surfaces include two small hole pockets $\alpha_1$ and $\alpha_2$ at the $Z$ point with dominant $d_{xz}/d_{yz}$ orbital characters, a large hole pocket $\gamma$ at the $R=(\pi,\pi,\pi)$ point of predominantly $d_{xy}$ orbital character, and two electron pockets $\beta_1$ $(\beta_2)$ situated at $X'=(\pi,0,\pi)$ $(Y'=(0,\pi,\pi))$ points of $d_{yz}/d_{xy}$ ($d_{xz}/d_{xy}$) orbital characters. Since the $Y'$-centered pocket is symmetry-related to the $X'$ pocket in this tetragonal system, it will not be discussed separately. In Fig.\ref{orbwt}, we show the corresponding orbital weight $|a^t_{\mu}(\mathbf{k})|^2$ for the $\beta_1$ pocket as a function of the angle $\phi$. We will relate this to $Z(\mathbf{k_\mu})$ and the modulation in gap amplitude in the following analysis of our results. 

Now, we will present our solutions to Eqs.~(\ref{Gapeqn}-\ref{Zeqn}) for $Z(\mathbf{k_\mu})$ and the leading pairing eigenvectors (gap functions). We have evaluated the results for two sets of Hubbard-Hund parameters ($U=0.7eV$, $J=0.26eV$) and ($U=0.79eV$, $J=0.2eV$). These parameters are close to the standard parameters used in the literature employing the RPA approach to the pairing vertex. 

First, we will discuss the results for $U=0.7eV$, $J=0.26eV$ and compare  with experimental data. In Fig.\ref{VA_HSP}, the diagonal components of the static particle-particle effective interaction $[V_{pppp}(\mathbf{q},0)]_A$ evaluated from Eq.~(\ref{VAint}) is plotted along the high-symmetry path $Z-R-X'-Z-Y'$. The RPA susceptibility shows an enhanced incommensurate peak around $\mathbf{q} = \pi(1,0.15)$ [similar to the findings in Ref.[\onlinecite{Wang13}]]. The $d_{xy}$ orbital channel is the largest in magnitude as a consequence of favorable nesting condition between $d_{xy}$-dominated parts of the electron pocket at $X'/Y'$ and the hole pocket at $R$. The phase space for scattering of quasiparticles between the $d_{xz}/d_{yz}$-dominated parts of the Fermi surfaces is restricted due to small pocket sizes of the $\alpha$ hole pockets and their unfavorable nesting conditions in connection to the electron pockets. Thus, the magnitude of the $d_{xz}/d_{yz}$ orbital channel for $[V_{pppp}(\mathbf{q},0)]_A$ is much lower than the $d_{xy}$ channel. Both the static particle-particle effective interaction $[V_{pqst}(\mathbf{q},0)]_A$ and particle-hole effective interaction $[V_{pqst}(\mathbf{q},0)]_N$ is dominated by the RPA spin-susceptibility contribution, and hence their underlying momentum structure is similar.

\begin{figure}[hbt!]
     \centering 
     \includegraphics[width=0.4\textwidth]{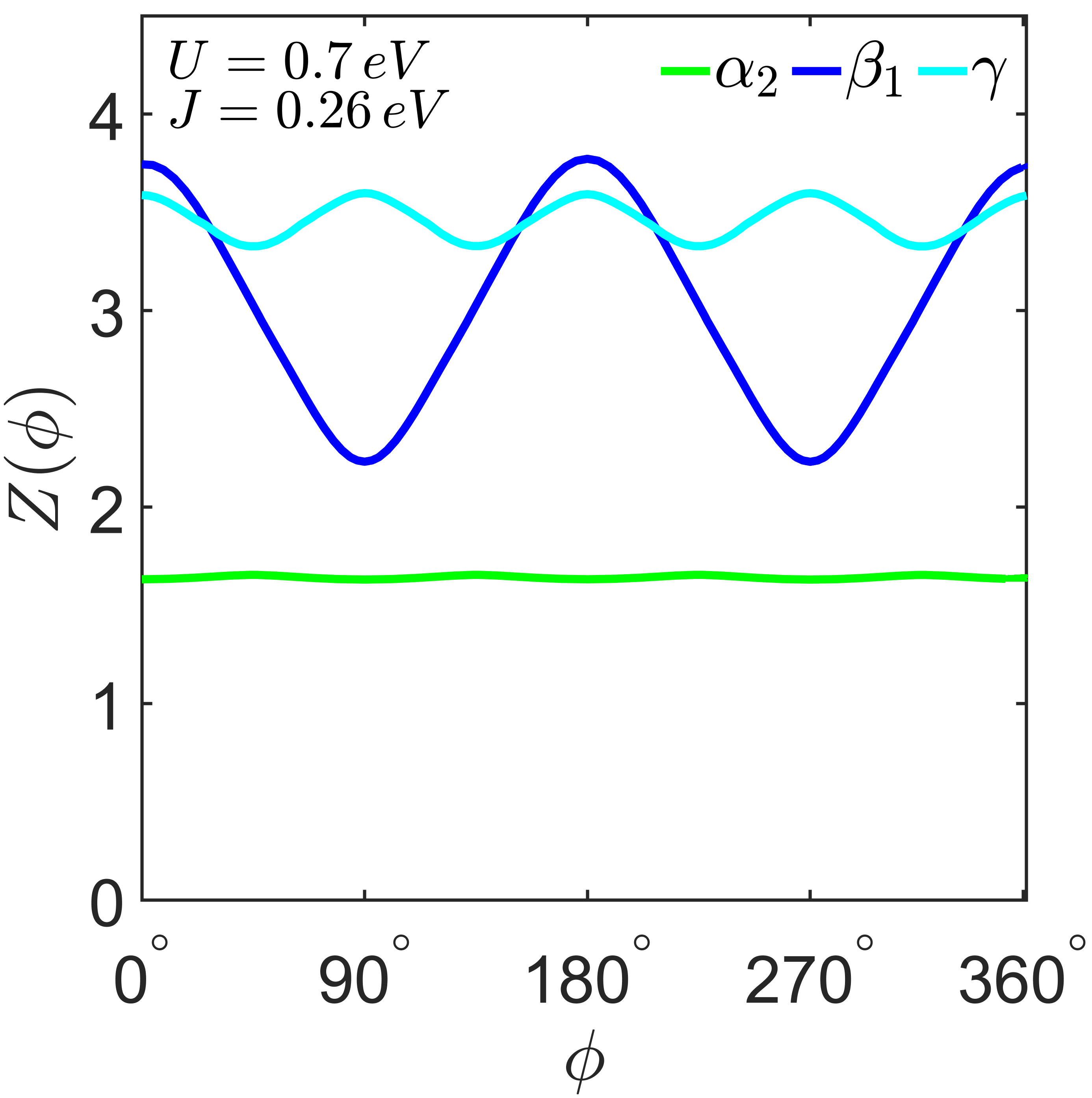}
     \caption{$Z(\phi)$ along Fermi pockets $\alpha_2, \beta_1$ and $\gamma$ evaluated from the static particle-hole effective interaction defined in Eq.~(\ref{Zeqn})}
     \label{Zkfig}
\end{figure}

In Fig.\ref{Zkfig}, we plot  $Z (\mathbf{k_\mu})$ evaluated from Eq.~(\ref{Zeqn}) as a function of the angle $\phi$ around the $\alpha_2,\beta_1$ and $\gamma$ Fermi pockets. Evidently, $Z (\phi)$ around the $\beta_1$ pocket follows the orbital content of the $d_{xy}$ orbital [compare to Fig.\ref{orbwt} for an angular plot of the orbital weight around the $\beta_1$ pocket]. This is expected from the current Fermi surface topology depending on two factors: (1) the interaction vertices are dominated by intraorbital processes, and (2) the $d_{xy}$ orbital with large weight at certain positions on the Fermi surface can take advantage of the strong peak in the susceptibility for scattering vectors $ \mathbf{k_\mu - k'_\nu} \approx \pi(1,0.15)$ roughly connecting $\mathbf{k_\mu}$ and $\mathbf{k'_\nu}$ across Fermi pockets. For the $\beta_1$ pocket, this happens around [$0^\circ,180^\circ$]. For the $\gamma$ pocket, maxima in $Z(\phi)$ are seen at [$0^\circ, 90^\circ, 180^\circ,270^\circ$]. Due to the lower magnitude of the $[V_{pppp}(\mathbf{q},0)]_N$ vertex for the $d_{xz}/d_{yz}$ orbitals, the magnitude of $Z(\phi)$ around the $\alpha_2$ pocket is smaller than around the $\beta_1$ and $\gamma$ pockets.

\subsection{Gap structure} \label{gapresults}

\begin{figure*}[hbt!]
    \centering
    \subfloat[]{\includegraphics[width=0.332\textwidth]{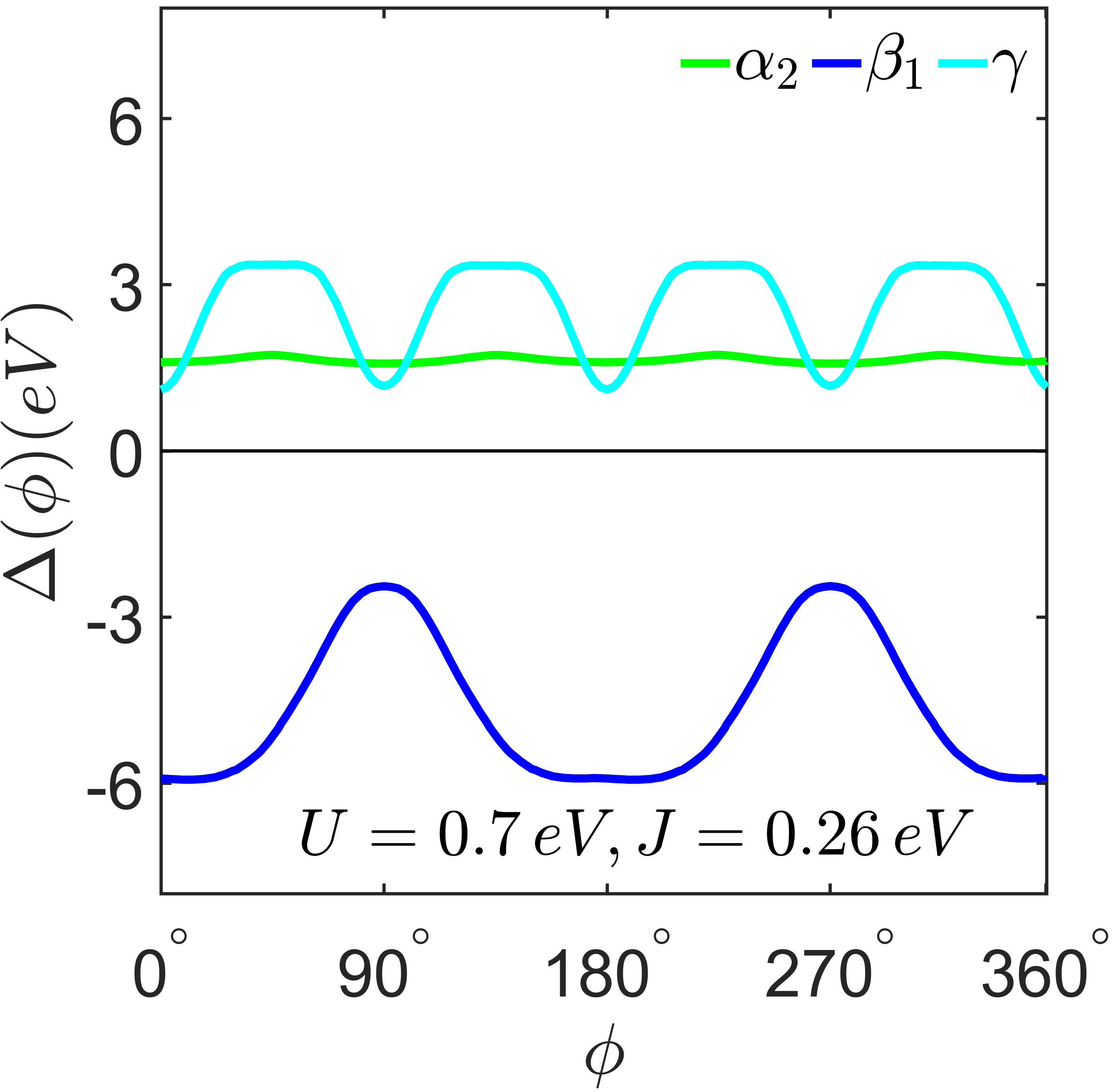}\label{oldgapplot}}
    \subfloat[]{\includegraphics[width=0.332\textwidth]{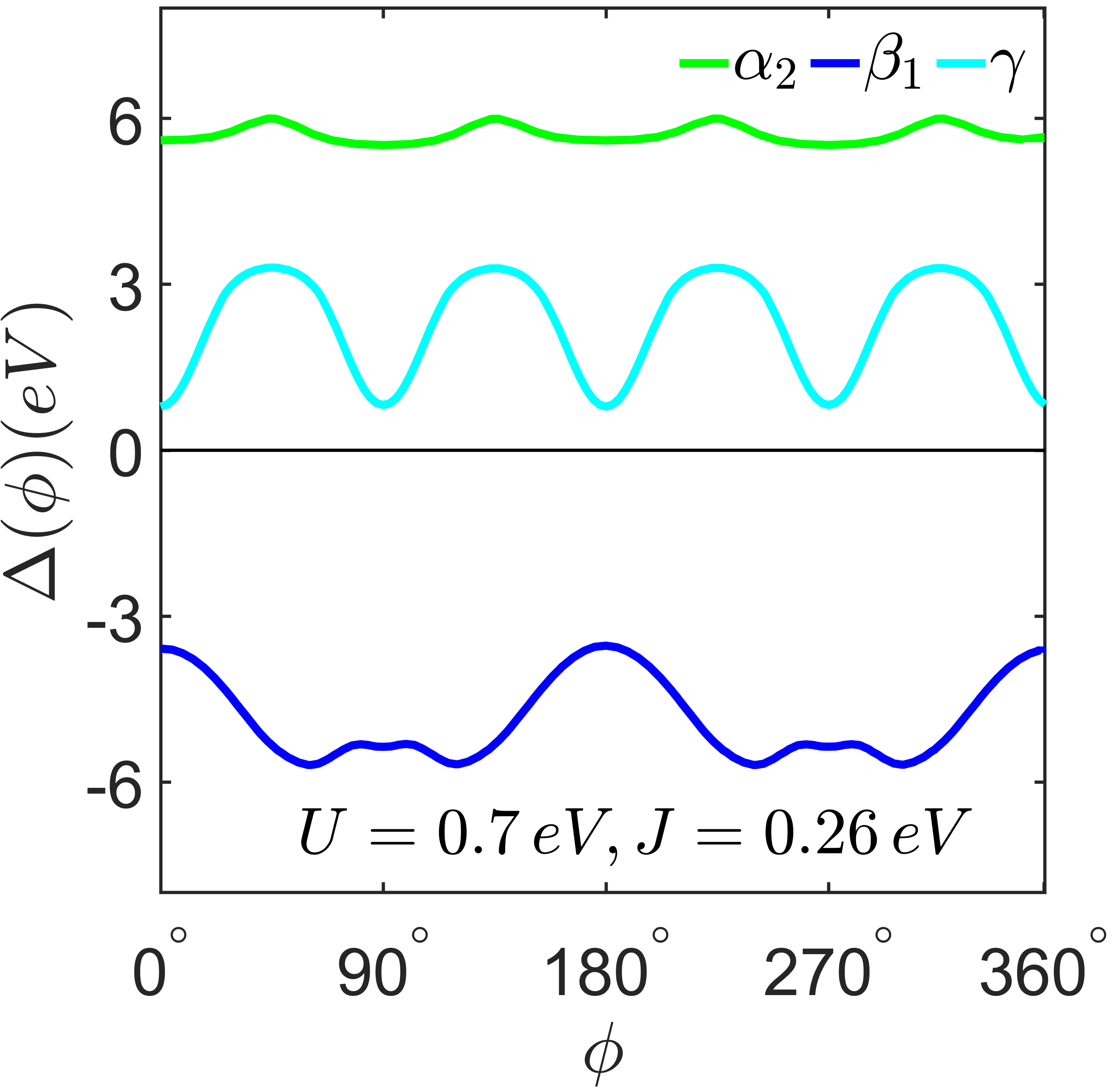}\label{newgapplot}}
    \subfloat[]{\includegraphics[width=0.333\textwidth]{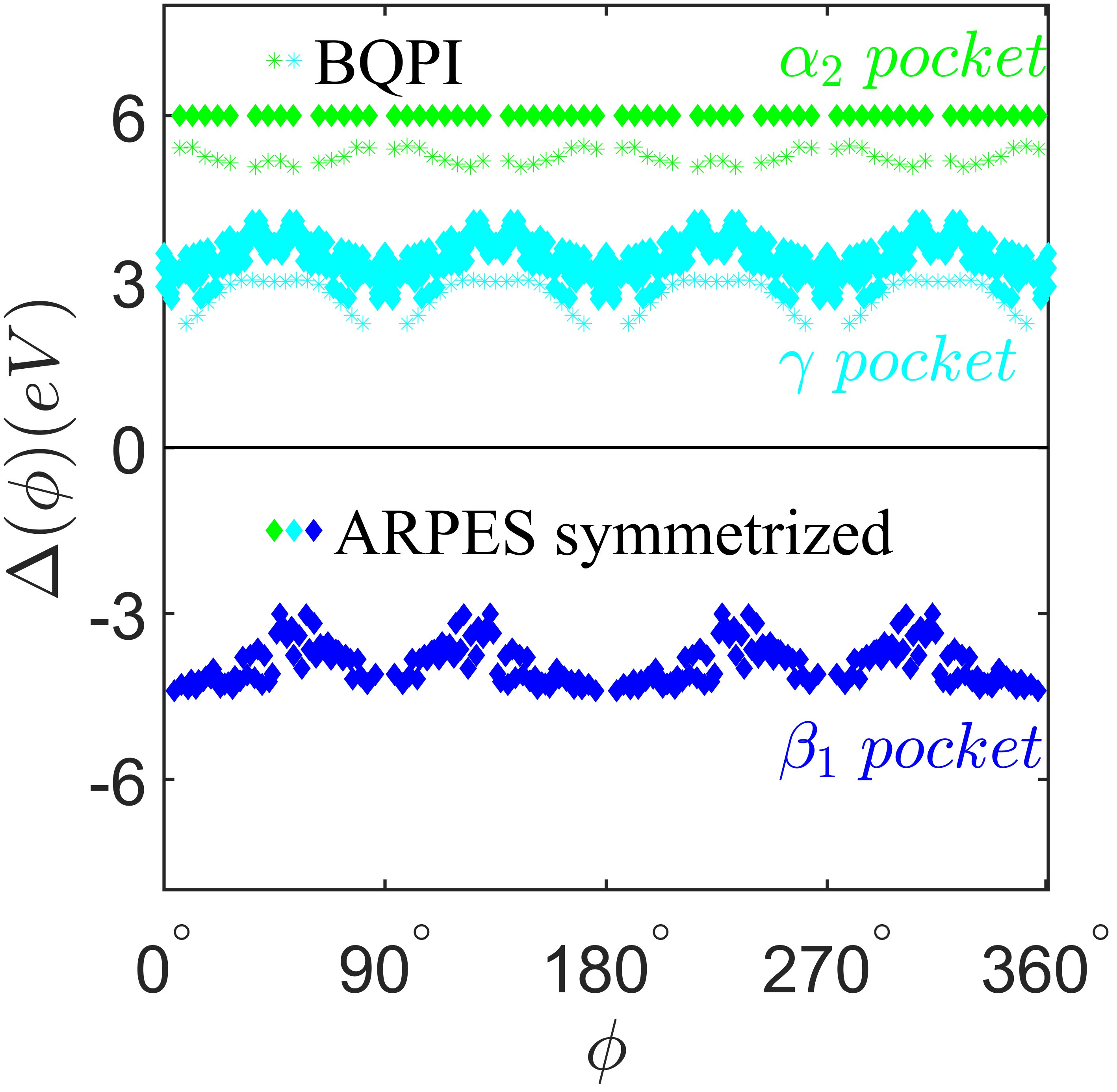}\label{expgapplot}}
    \caption{Results for LiFeAs: (a) Angular plot of the s-wave gap function evaluated from traditional spin-fluctuation theory, around the $Z$-centered outer hole pocket $\alpha_2$, $X'$-centered electron pocket $\beta_1$ and $R$-centered hole pocket $\gamma$. (b) The modified gap function obtained from the linearized Eliashberg equations~ (\ref{Gapeqn}-\ref{Zeqn}) with quasiparticle renormalization $Z({\bf k})$ included. (c) Experimental results: Measured magnitudes of the gap from an ARPES experiment [\onlinecite{Borisenko12}], symmetrized and displayed as diamonds and those from Bogoliubov QPI experiment [\onlinecite{Allan12}] displayed as crosses. All calculations are done for $U$ and $J$ as indicated.}
\end{figure*}

We show the results for the leading gap structure in Fig.~\ref{oldgapplot} obtained from traditional spin-fluctuation calculation with $Z(\mathbf{k'_\nu})$ set to unity in Eq.~(\ref{Gapeqn}). Fig.~\ref{newgapplot} is the most important result of this work, displaying the modified gap structure obtained by inclusion of momentum-dependent correlation effects $Z(\mathbf{k_\mu}) \neq 1 $ via the linearized Eliashberg equations~(\ref{Gapeqn}-\ref{Zeqn}). For comparison of our calculations to experimental results, we have plotted $C_4$-symmetrized ARPES data for the gap magnitude taken from Ref.~[\onlinecite{Borisenko12}] and BQPI data from Ref.~[\onlinecite{Allan12}] for the three Fermi pockets $ \alpha_2,\beta_1,\gamma$ in Fig.~\ref{expgapplot}. ARPES sees only one band crossing at the Fermi level at $Z$ on the $k_z=\pi$ plane with a large gap of the order of $6$ meV. However, our current DFT-derived tight-binding structure always produces two $\alpha$ pockets. Hence, we will consider it to be roughly appropriate to speak of an average gap on the $\alpha$ pockets assigned to $\alpha_2$. In both Figs.~\ref{oldgapplot} and \ref{newgapplot}, we find an $s^\pm$-wave state with highly anisotropic but full gaps on the electron (negative gap) and hole (positive gap) pockets. 

With self-energy effects included, the gap functions undergo a remarkable change [Fig.~\ref{newgapplot}] relative to the traditional spin-fluctuation calculation [Fig.~\ref{oldgapplot}]. First, we find a stronger tendency towards $s^\pm$ pairing symmetry, even for small values of $J$. We will discuss this in further detail in Section~\ref{spm_symmetry}. Second,  $Z(\mathbf{k_\mu})$ induces a momentum-dependent modulation of the gap function on various pockets. Most importantly, in Fig.~\ref{newgapplot} we see that $Z(\mathbf{k_\mu})$ enhances the magnitude of the nearly-isotropic gap on the small $Z$-centered hole pocket $\alpha_2$. This brings the results of spin-fluctuation theory much closer in line with experimental data, and corrects the crucial discrepancy in the calculation of Wang \textit{et al.} [\onlinecite{Wang13}] relative to experiment. Although the angular positions of the gap maxima and minima on the $\gamma$ pocket and the average gap magnitude is in agreement with experiment, we do not obtain a suppression of the anisotropy. We find weaker anisotropy of the gap function on the $\beta_1$ pocket and an average gap magnitude comparable to experiment. However, the theoretical gap structure has 2 global maxima as opposed to 4 in the experimental data, and these maxima are located at positions where the experimental data has minima. In the next section we will analyze the results for the gap structure in terms of the quasiparticle renormalization factor $Z(\mathbf{k_\mu})$ and the pairing vertex.

\subsection{Analysis of the gap structure in terms of quasiparticle weights and the pairing vertex} \label{gapanalysis}

\begin{figure*}[hbt!]
    \centering
    \subfloat[]{\includegraphics[width=0.326\textwidth]{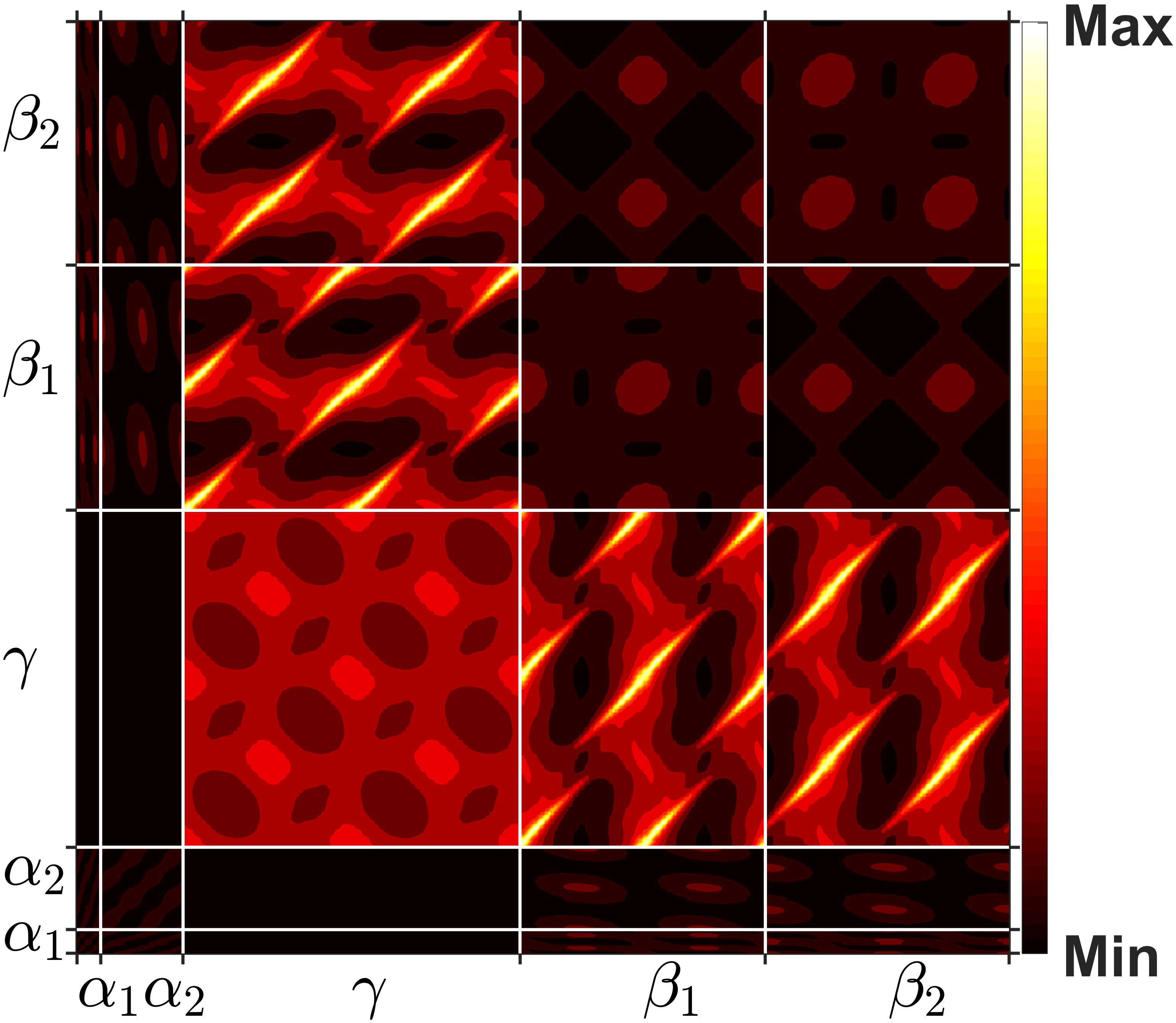}\label{ori_gsij}} \,
    \subfloat[]{\includegraphics[width=0.326\textwidth]{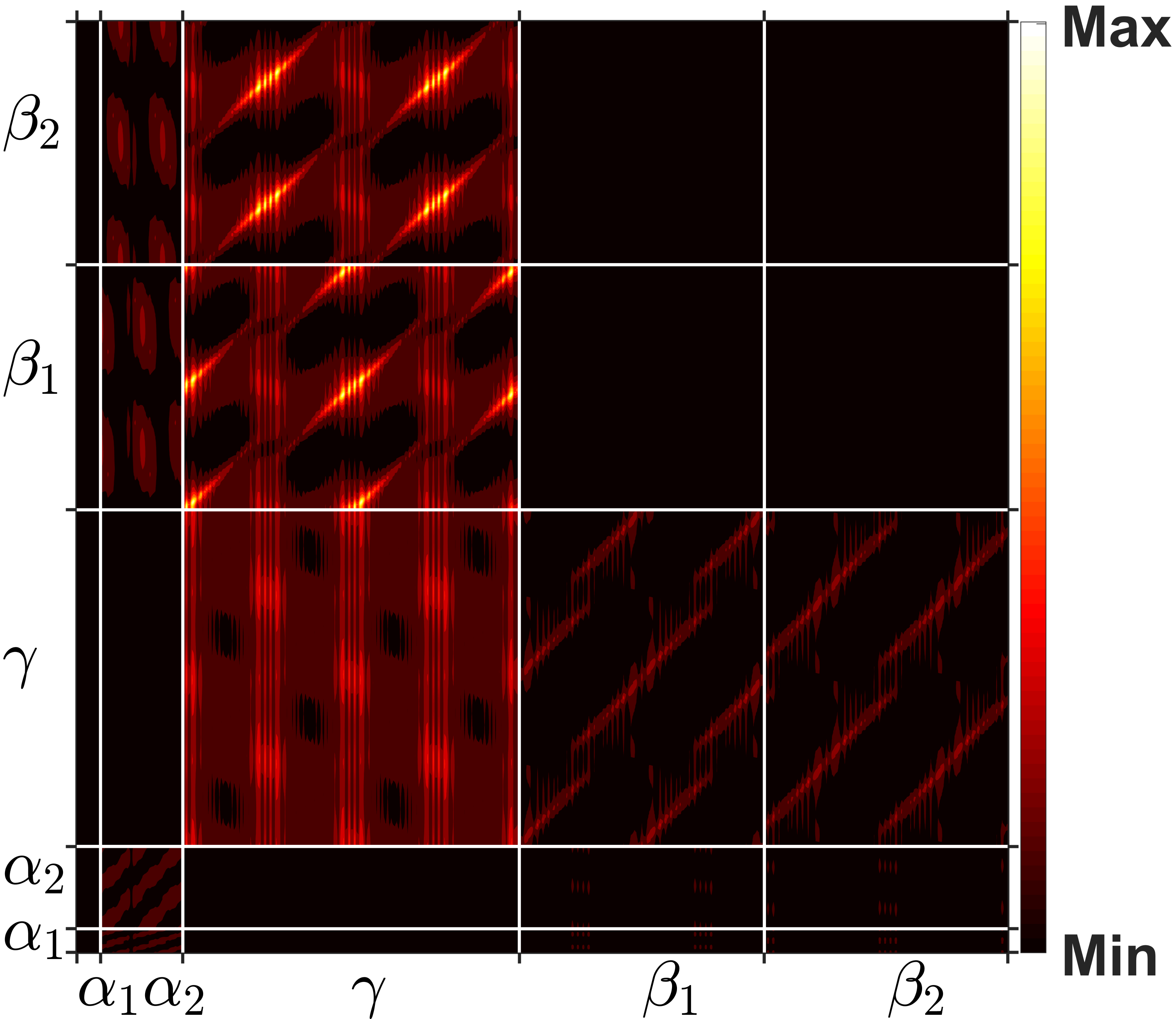}\label{gsij_unsymm}} \,
    \subfloat[]{\includegraphics[width=0.326\textwidth]{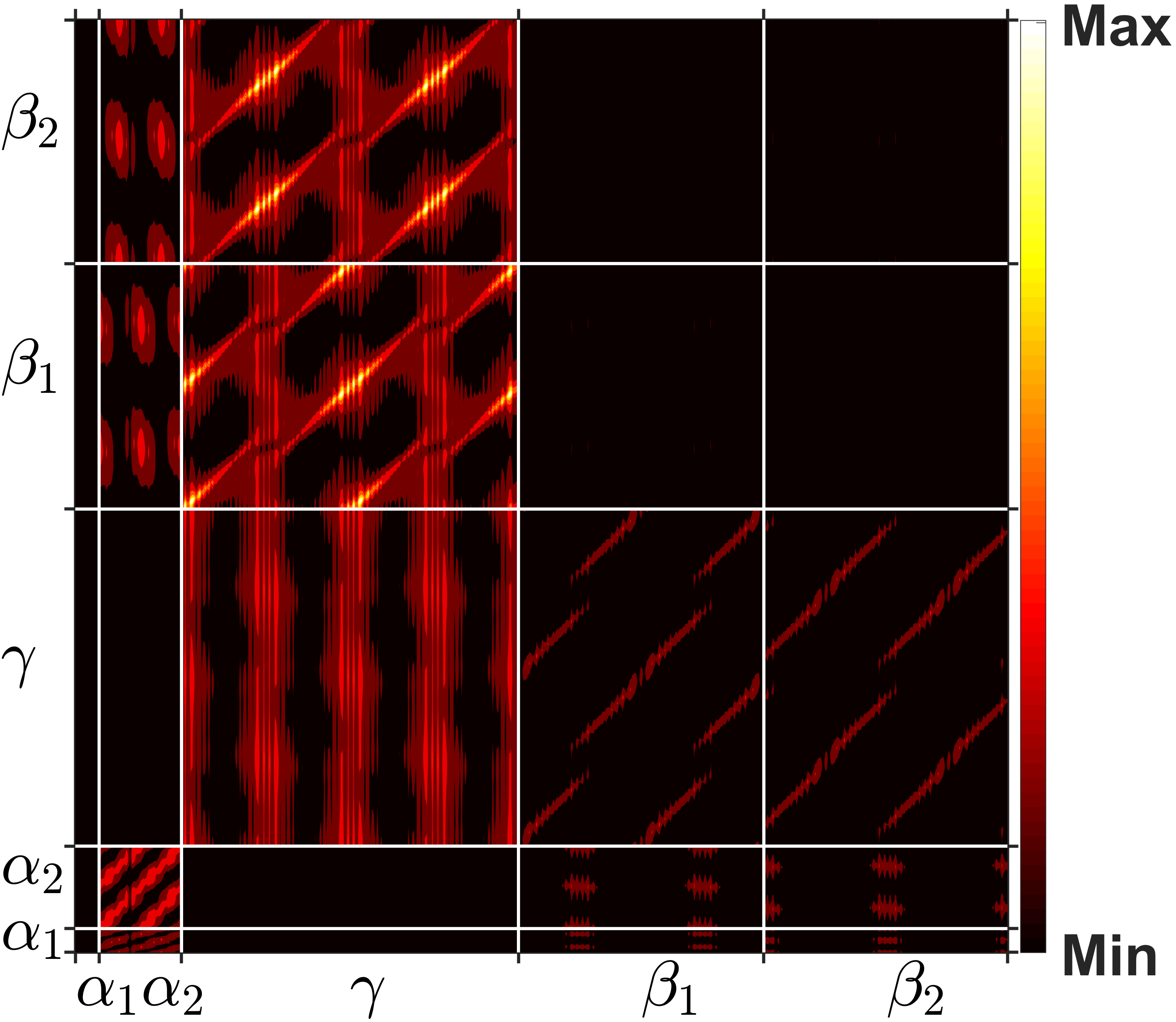}\label{gsij_Zk_unsymm}}
    \caption{Heatmap plots of (a) the singlet pairing vertex $ \left[ V(\mathbf{k_\mu,k'_\nu},0) \right]_A $ matrix resulting in the gap function plotted in Fig.~\ref{oldgapplot} from the tight-binding model, (b) Matrix elements of the quantity $ \left[ V(\mathbf{k_\mu,k'_\nu},0) \right]_A dk'_\nu/|\mathbf{v}_F(\mathbf{k'_\nu})| $, (c) Matrix elements of the quantity $ \left[ V(\mathbf{k_\mu,k'_\nu},0) \right]_A dk'_\nu/|\mathbf{v}_F(\mathbf{k'_\nu})| / Z(\mathbf{k'_\nu}) $. The value at any given point is proportional to the brightness of the color. The rows and the columns of the tiles of (a)-(c) correspond to the Fermi points $\mathbf{k_\mu}$ and $\mathbf{k'_\nu}$ arranged in ascending order of their corresponding polar angle $\phi = 0^\circ$ to $360^\circ$ around each Fermi sheet. Band indices $\mu $ and $\nu$ represent the Fermi sheets $\alpha_1,\alpha_2$ at the $Z$, $\gamma$ at the $R$, $\beta_1$ at the $X'$ and $\beta_2$ at the $Y'$ point. Unequal areas of the tiles are due to unequal number of Fermi points around each of the sheets.} 
\end{figure*}

We investigate the structure of the effective pair vertex from its graphical representation in Fig.~\ref{ori_gsij}-\ref{gsij_Zk_unsymm}. Each block demarcated by white lines in the image represents a matrix $(\mathbf{k_\mu,k'_\nu})$ consisting of values corresponding to the pairing vertex, with the Fermi points $\mathbf{k_\mu}$ and $\mathbf{k'_\nu}$ arranged in ascending order of their corresponding polar angle $\phi = 0^\circ$ to $360^\circ$ around each Fermi sheet ($\alpha_1,\alpha_2,\beta_1,\beta_2,\gamma$). By construction, as in Eq.~(\ref{singletvertex}), $\left[ V(\mathbf{k_\mu,k'_\nu},0) \right]_A$ is a symmetric matrix with respect to the interchange $\mathbf{k_\mu} \longleftrightarrow \mathbf{k'_\nu}$ as seen in Fig.~\ref{ori_gsij}. Fig.~\ref{gsij_unsymm} shows matrix elements of the quantity $ \left[ V(\mathbf{k_\mu,k'_\nu},0) \right]_A dk'_\nu/|\mathbf{v}_F(\mathbf{k'_\nu})| $. This is the matrix which is diagonalized in traditional spin-fluctuation theory to obtain the leading eigenstate. It is not symmetric under the exchange $\mathbf{k_\mu} \longleftrightarrow \mathbf{k'_\nu}$. Here, $dk'_\nu$ is the length element of the Fermi sheet at the $k'_\nu$-th Fermi point. The factor $ 1/|\mathbf{v}_F(\mathbf{k'_\nu})| $ acts as a momentum-dependent density of states for the corresponding band $\nu$ at the Fermi level. 

In the figures, the brightest set of blocks are the ones representing scattering processes among the pockets  $\gamma$ and $\beta_1 \, (\beta_2)$. It is visually prominent that the dominant scattering processes occur for $\gamma\rightarrow \beta_1 \, (\beta_2) $. As discussed earlier, scattering between the $d_{xy}$-dominated part of the Fermi pockets contributes to the primary pairing interaction leading to superconductivity. Although angular positions of [$0^\circ,90^\circ,180^\circ,270^\circ$] on the $\gamma$ pocket are favored to take advantage of the $d_{xy}$-dominated scattering to either $\beta_1$ or $\beta_2$ pocket, positions of [$45^\circ,135^\circ,225^\circ,315^\circ$] take advantage of simultaneous scattering to $\beta_1$ and $\beta_2$ pockets. This renders gap maxima around [$45^\circ,135^\circ,225^\circ,315^\circ$] on the $\gamma$ pocket [refer to Fig.~\ref{oldgapplot}]. It also explains why gap maxima occur at [$0^\circ,180^\circ$] on the $\beta_1$ pocket due to favorable $d_{xy}$ scattering to the $\gamma$ pocket. Due to restricted scattering between $\alpha_2$ and the other pockets, the overall gap magnitude is very low on this pocket.   

In Fig.~\ref{gsij_Zk_unsymm}, we show matrix elements of the quantity $ \left[ V(\mathbf{k_\mu,k'_\nu},0) \right]_A dk'_\nu/|\mathbf{v}_F(\mathbf{k'_\nu})| / Z(\mathbf{k'_\nu}) $. The leading eigenfunction $ g(\mathbf{k_\mu}) $ of this matrix gives the modified gap function as $ \Delta(\mathbf{k_\mu}) = \frac{ g(\mathbf{k_\mu}) }{ Z(\mathbf{k_\mu}) }$. Compared to Fig.~\ref{gsij_unsymm}, the first noticeable difference is the enhanced pairing (brighter spots) occurring between the $\alpha_2$ pocket and the $\beta_1 \, (\beta_2)$ pockets. It appears as if the electronic correlations tend to lift the restriction on the phase space of scattering between the non-nested $d_{xz}/d_{yz}$-dominated parts of the $\alpha_2$ Fermi pocket. This is one of the main effects of the momentum-dependent modulation of the pairing vertex due to $Z(\mathbf{k'_\nu})$. In addition, due to the smaller magnitude of $Z(\mathbf{k_{\alpha_2}})$ [refer to Fig.~\ref{Zkfig}], the final gap function $ \Delta(\mathbf{k_{\alpha_2}}) $ undergoes further enhancement in its magnitude as seen in Fig.~\ref{newgapplot}. We also find that the gap maxima shifts to [$90^\circ,270^\circ$] on the $\beta_1$ pocket due to enhanced pairing to the $\alpha_2$ pocket via the $d_{yz}$ orbitals, combined with the effect of momentum modulation by $Z(\mathbf{k_{\beta_1}})$ on the final gap function. The $\gamma$ and $\beta_1 \, (\beta_2) $ pocket pairing is not affected substantially due to the $Z(\mathbf{k'_\nu})$ modulation. Hence, $ \Delta(\mathbf{k_{\gamma}}) $ does not undergo any drastic changes.

\subsection{Discussion: Tendency towards \texorpdfstring{$s\pm$}{} pairing symmetry} \label{spm_symmetry}

\begin{figure}[hbt!]
    \centering
    \includegraphics[width=\linewidth]{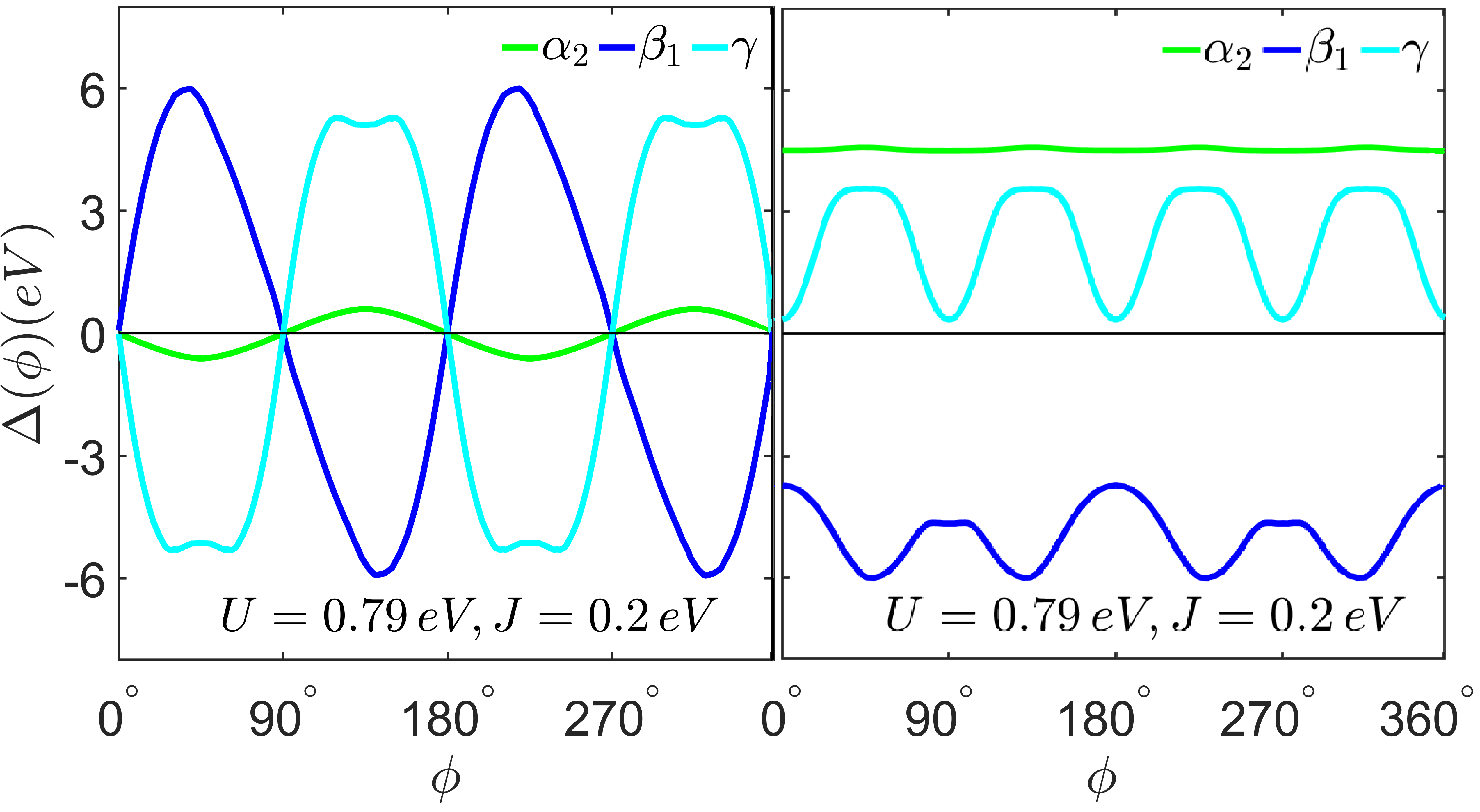} 
    \caption{Angular plots of the gap function evaluated around the Fermi pockets as indicated by the color legend. The calculations show a change in the gap symmetry of the leading eigenfunction from (left) $d$-wave (obtained from traditional spin-fluctuation theory with $Z(\bf k)=1$) to (right) $s$-wave (obtained from linearized Eliashberg equations with $Z(\bf k)$ included).  Both calculations are done for $U$ and $J$ as indicated.}
    \label{changesymmplot}
\end{figure}

A close competitor of the leading $s^\pm$ state is the $d$-wave state, which has never been convincingly observed in any experiment in the context of FeSCs. In the traditional spin-fluctuation scenario, $d$ and $s$ wave solutions are nearly degenerate in LiFeAs  [\onlinecite{Wang13}],  a consequence of its poor $(\pi,0)$ nesting properties [\onlinecite{Maiti2011},\onlinecite{Lee_etal_Kotliar2012},\onlinecite{BorisenkoLiFeAs}]. In Fig.~\ref{changesymmplot}, we show the results for calculations with Hubbard-Hund parameters ($U=0.79eV$, $J=0.2eV$) [which is a higher $U$ and a lower $J$ value than the previous calculation shown in Fig.\ref{oldgapplot}-\ref{newgapplot}]. With this particular set of $U$ and $J$, the traditional spin-fluctuation theory yields a $d$-wave state as the leading pairing state (left panel in Fig.~\ref{changesymmplot}). We see a sign change in the gap structure around each pocket ($\alpha_2, \beta_1, \gamma$) with a periodicity of $\pi/2$. For this case, the second leading eigen channel is the $s$-wave channel and the ratio of their eigen values is $\lambda_d/\lambda_s = 1.05$.  In contrast, the inclusion of correlations effect via the linearized Eliashberg equations \ref{Gapeqn}-\ref{Zeqn} stabilizes the $s$-wave state as the leading eigenfunction, and the $d$-wave state becomes subleading. For this case, we get $\lambda_d/\lambda_s = 0.97$. This points towards quasiparticle renormalization effects playing an important role in stabilizing the $s^\pm$ pairing state in LiFeAs. 

\section{Conclusions}

We have performed 2D calculations of the superconducting pairing state for the LiFeAs compound, one of the few materials where ARPES experiments have measured the gap function on several different Fermi surface sheets, and found significant gap anisotropy.  As a test of the proposal that a proper treatment of momentum dependent quasiparticle renormalization is the major missing ingredient in traditional spin fluctuation calculations of gap structures in FeSC, we have provided a simple approximate theoretical framework for the calculation of the superconducting gap from spin fluctuation theory, including momentum-dependent self-energy effects at the single-particle level. We have shown that this approach leads to robust $s^\pm$ pairing of the conventional type, and  provides  substantial  improvement  of the calculated gap structure on all the Fermi pockets compared to experiments, especially for the large gap on the inner hole pocket. We have calculated the leading stationary solution of a spin-fluctuation pairing strength functional which includes quasiparticle self-energy effects, and discussed their role in stabilizing the $s^\pm$ eigenstate. We conclude that the inclusion of self-energy effects within the framework of existing weak-coupling theories is an important ingredient to understand the observed superconducting pairing structures in LiFeAs, and other FeSCs, which are dictated by the same underlying physical processes. Vertex corrections as calculated in Ref.[\onlinecite{Saito14}] do not appear to be necessary to capture the basic renormalization of the electronic structure and pairing vertex.

\section*{Acknowledgments}

We acknowledge useful discussions with A.Kreisel, S. Maiti, C. Setty, and L. Fanfarillo. S.B. acknowledges support in part through an appointment to the Oak Ridge National Laboratory ASTRO Program, sponsored by the U.S. Department of Energy and administered by the Oak Ridge Institute for Science and Education. P.J.H.  was supported under the Grant No. DE-FG02-05ER46236. T.A.M. was supported by the U.S. Department of Energy, Office of Basic Energy Sciences, Materials Sciences and Engineering Division. The work of D.J.S. was supported by the Scientific Discovery through Advanced Computing (SciDAC) program funded by U.S. Department of Energy, Office of Science, Advanced Scientific Computing Research and Basic Energy Sciences, Division of Materials Sciences and Engineering.

\onecolumngrid
\appendix
\section{A spin-fluctuation pairing strength functional including quasiparticle renormalization } \label{appendics}

In the multiband Nambu basis given by:
\[ \Psi_{\alpha \mathbf{k}} = \left( \begin{matrix}
    c_{\alpha, \mathbf{k}\uparrow} \\
    c^\dagger_{\alpha,-\mathbf{k}\downarrow} \\
    \end{matrix} \right)  
, \, \Psi_{\alpha \mathbf{k}}^\dagger = \left( \begin{matrix}
    c^\dagger_{\alpha,\mathbf{k}\uparrow} \, , \, c_{\alpha,-\mathbf{k}\downarrow} \\
    \end{matrix} \right) \]
the band index $\alpha$ ranges from $1$ to $5$ in FeSCs equal to the number of Fe $3d$ orbitals. We express the non-interacting Hamiltonian as $ H_0 = \displaystyle{\sum_{\alpha \mathbf{k} } } \varepsilon_\alpha(\mathbf{k}) \Psi^\dagger_{\alpha \mathbf{k}} \tau_3 \Psi_{\alpha \mathbf{k}} $ where $\varepsilon_\alpha(\mathbf{k})$ is the dispersion in the $\alpha$ band in the normal state of the system. The Pauli matrices are denoted conventionally by $\tau_i, i=0,1,2,3$. The Hamiltonian $H_0$, the full renormalized Green's function $\underline{G}(\mathbf{k},\omega_m)$, the non-interacting Green's function  $\underline{G}_0(\mathbf{k},\omega_m)$, and the self-energy $\underline{\Sigma}(\mathbf{k},\omega_m)$ are all $10 \times 10$ matrices in this multiband basis for every set of $(\mathbf{k},\omega_m)$. 
The non-interacting band Green's function $G_{\alpha 0}(\mathbf{k},\omega_m) = (i\omega_m- \varepsilon_\alpha(\mathbf{k}))^{-1}$ gives the full matrix as $\underline{G}_0(\mathbf{k},\omega_m) = (i\omega_m \tau_0 - \varepsilon_\alpha(\mathbf{k}) \tau_3)^{-1}$. The particle-hole and the particle-particle Green's function are defined, respectively, as:
\begin{align}
\begin{split}
    G_\alpha(\mathbf{k},\omega_m) &= - \int_0^{\beta} d\tau e^{i\omega_m \tau} \langle T_\tau c_{\alpha,\mathbf{k}\sigma}(\tau) c^\dagger_{\alpha,\textbf{k} \sigma}(0)\rangle \\
  F_\alpha(\mathbf{k},\omega_m) &= \int_0^{\beta} d\tau e^{i\omega_m \tau} \langle T_\tau c_{\alpha,\mathbf{k}\uparrow}(\tau) c_{\alpha,-\textbf{k} \downarrow}(0)\rangle \\
\end{split}        
\end{align}
and the corresponding full Green's function:
\[
\underline{G}(\mathbf{k},\omega_m) = \left( \begin{matrix}
    G_\alpha(\mathbf{k},\omega_m) & F_\alpha(\mathbf{k},\omega_m) \\
    F_\alpha^*(\mathbf{k},\omega_m) & -G_\alpha(\mathbf{-k},-\omega_m) \\
    \end{matrix} \right) 
\]
The full self-energy is:
\begin{equation}
    \underline{\Sigma}(\mathbf{k},\omega_m) = \left( \begin{matrix}
    \Sigma_\alpha(\mathbf{k},\omega_m) & \Phi_\alpha(\mathbf{k},\omega_m) \\
    \Phi_\alpha^*(\mathbf{k},\omega_m) & -\Sigma_\alpha(\mathbf{-k},-\omega_m) \\
    \end{matrix} \right)  
\end{equation}
where the particle-hole and particle-particle components are given by:
\begin{equation}
\begin{split}
    \label{self_particle-particle}
    \Sigma_\alpha(\mathbf{k},\omega_m) &= \frac{1}{\beta N_{k'}} \displaystyle{\sum_{n,\beta,\mathbf{k'}}} V_{\alpha\beta N}(\mathbf{k-k'},\omega_m-\omega_n) G_\beta(\mathbf{k'},\omega_n) \\
    \Phi_\alpha(\mathbf{k},\omega_m) &= \frac{1}{\beta N_{k'}} \displaystyle{\sum_{n,\beta,\mathbf{k'}}} V_{\alpha\beta A}(\mathbf{k-k'},\omega_m-\omega_n) F_\beta(\mathbf{k'},\omega_n)
\end{split}
\end{equation}
The particle-hole interaction between band $\alpha$ and $\beta$ is denoted by $V_{\alpha\beta N}(\mathbf{k-k'},\omega_m-\omega_n)$ whereas the particle-particle interaction is denoted by $V_{\alpha\beta A}(\mathbf{k-k'},\omega_m-\omega_n)$.  
From Dyson's equation, we have: 
\begin{equation}
\label{Dyson}
    \underline{G}^{-1}(\mathbf{k},\omega_m) = \underline{G}^{-1}_0(\mathbf{k},\omega_m) - \underline{\Sigma}(\mathbf{k},\omega_m)
\end{equation}
Using the Pauli matrices, $\underline{\Sigma}$ can be written as:
\begin{equation}
\label{self-expansion}
    \underline{\Sigma}(\mathbf{k},\omega_m) = i\omega_m [1-Z_\alpha(\mathbf{k})]\tau_0 + X_\alpha(\mathbf{k},\omega_m)\tau_3 + \Phi_\alpha(\mathbf{k},\omega_m)\tau_1 + \overline{\Phi_\alpha}(\mathbf{k},\omega_m)\tau_2 
\end{equation}
with yet unknown and independent functions $Z_\alpha(\mathbf{k}), X_\alpha(\mathbf{k},\omega_m), \Phi_\alpha(\mathbf{k},\omega_m), \overline{\Phi_\alpha}(\mathbf{k},\omega_m)$. From Dyson's Eq.\ref{Dyson} and the Pauli matrix identity $ (a_0\tau_0 + \Vec{a}.\Vec{\tau})(a_0\tau_0 - \Vec{a}.\Vec{\tau}) = (a_0^2 -\Vec{a}^2)\tau_0$, one finds:
\begin{equation}
    \underline{G}(\mathbf{k},\omega_m) = \left[ i\omega_m Z_\alpha(\mathbf{k})\tau_0 + ( \varepsilon_\alpha(\mathbf{k}) + X_\alpha(\mathbf{k},\omega_m)) \tau_3 + \Phi_\alpha(\mathbf{k},\omega_m)\tau_1 + \overline{\Phi_\alpha}(\mathbf{k},\omega_m)\tau_2 \right]/ D_\alpha(\mathbf{k},\omega_m)
\end{equation}
with $ D_\alpha(\mathbf{k},\omega_m) = \text{det} \underline{G}^{-1}(\mathbf{k},\omega_m) = (i\omega_m Z_\alpha(\mathbf{k}))^2 - ( \varepsilon_\alpha(\mathbf{k}) + X_\alpha(\mathbf{k},\omega_m) )^2  - \Phi_\alpha(\mathbf{k},\omega_m)^2- \overline{\Phi_\alpha}(\mathbf{k},\omega_m)^2 $. Using Eq.\ref{self_particle-particle} and \ref{self-expansion}, we obtain the following self-consistent equations for the four unknown functions:

\begin{equation}
\begin{split}
        i\omega_m (1-Z_\alpha(\mathbf{k})) &= \frac{1}{\beta N_{k'}} \displaystyle{\sum_{n,\beta, \mathbf{k'}}} V_{\alpha\beta N}(\mathbf{k-k'},\omega_m-\omega_n) \frac{i\omega_{n} Z_\beta(\mathbf{k'})}{D_\beta(\mathbf{k'},\omega_{n})} \\
        X_\alpha(k,\omega_m) &= \frac{1}{\beta N_{k'}} \displaystyle{\sum_{n,\beta, \mathbf{k'}}} V_{\alpha\beta N}(\mathbf{k-k'},\omega_m-\omega_n) \frac{\varepsilon_\beta(\mathbf{k'}) + X_\beta(\mathbf{k'},\omega_n)}{D_\beta(\mathbf{k'},\omega_{n})} \\
        \Phi_\alpha(\mathbf{k},\omega_m) &= \frac{1}{\beta N_{k'}} \displaystyle{\sum_{n,\beta, \mathbf{k'}}} V_{\alpha\beta A}(\mathbf{k-k'},\omega_m-\omega_n) \frac{\Phi_\beta(\mathbf{k'},\omega_n)}{D_\beta(\mathbf{k'},\omega_{n})} \\
        \overline{\Phi_\alpha}(\mathbf{k},\omega_m) &= \frac{1}{\beta N_{k'}} \displaystyle{\sum_{n,\beta, \mathbf{k'}}} V_{\alpha\beta A}(\mathbf{k-k'},\omega_m-\omega_n) \frac{\overline{\Phi_\beta}(\mathbf{k'},\omega_n)}{D_\beta(\mathbf{k'},\omega_{n})} \\
\end{split}
\end{equation}
With the modified dispersion given by the poles of the Green's function, we define: 
\begin{equation}
\label{BDG}
E_\alpha(\mathbf{k},\omega_n) = \sqrt{ \frac{( \varepsilon_\alpha(\mathbf{k}) + X_\alpha(\mathbf{k},\omega_n))^2}{Z_\alpha(\mathbf{k})^2}  + \frac{\Phi_\alpha(\mathbf{k},\omega_n)^2 + \overline{\Phi_\alpha}(\mathbf{k},\omega_n)^2}{Z_\alpha(\mathbf{k})^2} } 
\end{equation}
The normal state corresponds to a solution $\Phi=\overline{\Phi}=0$. $Z$ is the quasiparticle renormalization factor, and $X$ describes shifts in the electron energies. The superconducting state is characterized
by a non-zero $\Phi$ or $\overline{\Phi}$. From Eq.\ref{BDG} the gap function given by:
\begin{equation}
\label{GAP}
\Delta_\alpha(\mathbf{k},\omega_n) = \frac{\Phi_\alpha(\mathbf{k},\omega_n) - i \overline{\Phi_\alpha}(\mathbf{k},\omega_n)}{Z_\alpha(\mathbf{k})} 
\end{equation}
describes the energy gap in the quasiparticle spectrum. $\Phi(\mathbf{k},\omega_n) $ and $ \overline{\Phi}(\mathbf{k},\omega_n) $ obey the same equations and are expected to have the same functional form up to a common phase factor. This phase factor becomes important in the description of Josephson junctions, but is irrelevant for the
thermodynamic properties of a homogeneous superconductor. In the following, we choose the simple gauge $\overline{\Phi}(\mathbf{k},\omega_n) = 0$, rendering $ \Delta_\alpha(\mathbf{k},\omega_n) = \frac{\Phi_\alpha(\mathbf{k},\omega_n)}{Z_\alpha(\mathbf{k})} $. We also rewrite $ D_\alpha(\mathbf{k},\omega_{n}) = (i\omega_{n} Z_\alpha(\mathbf{k}))^2 - ( E_\alpha(\mathbf{k},\omega_n) Z_\alpha(\mathbf{k}))^2 $. For simplifying the self-consistent equations further, we will ignore $X_\alpha(\mathbf{k},\omega_n)$ considering energy shifts to the dispersion being much smaller and negligible compared to the value of $\varepsilon_\alpha(\mathbf{k})$ itself. This leaves us with the following two equations to solve:
\begin{subequations}
\begin{align}
    \label{Eliash1}
        i\omega_m (1-Z_\alpha(\mathbf{k})) &= \frac{1}{\beta N_{k'}} \displaystyle{\sum_{n,\beta, \mathbf{k'}}} V_{\alpha\beta N}(\mathbf{k-k'},\omega_m - \omega_n) \frac{i\omega_{n} }{Z_\beta(\mathbf{k'}) \left( (i\omega_{n} )^2 -  E_\beta(\mathbf{k'},\omega_n)^2 \right)} \\
    \label{Eliash2}
        \Phi_\alpha(\mathbf{k},\omega_m) &= \frac{1}{\beta N_{k'}} \displaystyle{\sum_{n,\beta, \mathbf{k'}}} V_{\alpha\beta A}(\mathbf{k-k'},\omega_m-\omega_n) \frac{\Phi_\beta(\mathbf{k'},\omega_n)}{Z_\beta^2(\mathbf{k'}) \left( (i\omega_{n})^2 - E_\beta(\mathbf{k'},\omega_n)^2 \right) } 
\end{align}
\end{subequations}
\begin{figure}[hbt!]
    \centering
    \includegraphics[width=0.4\textwidth]{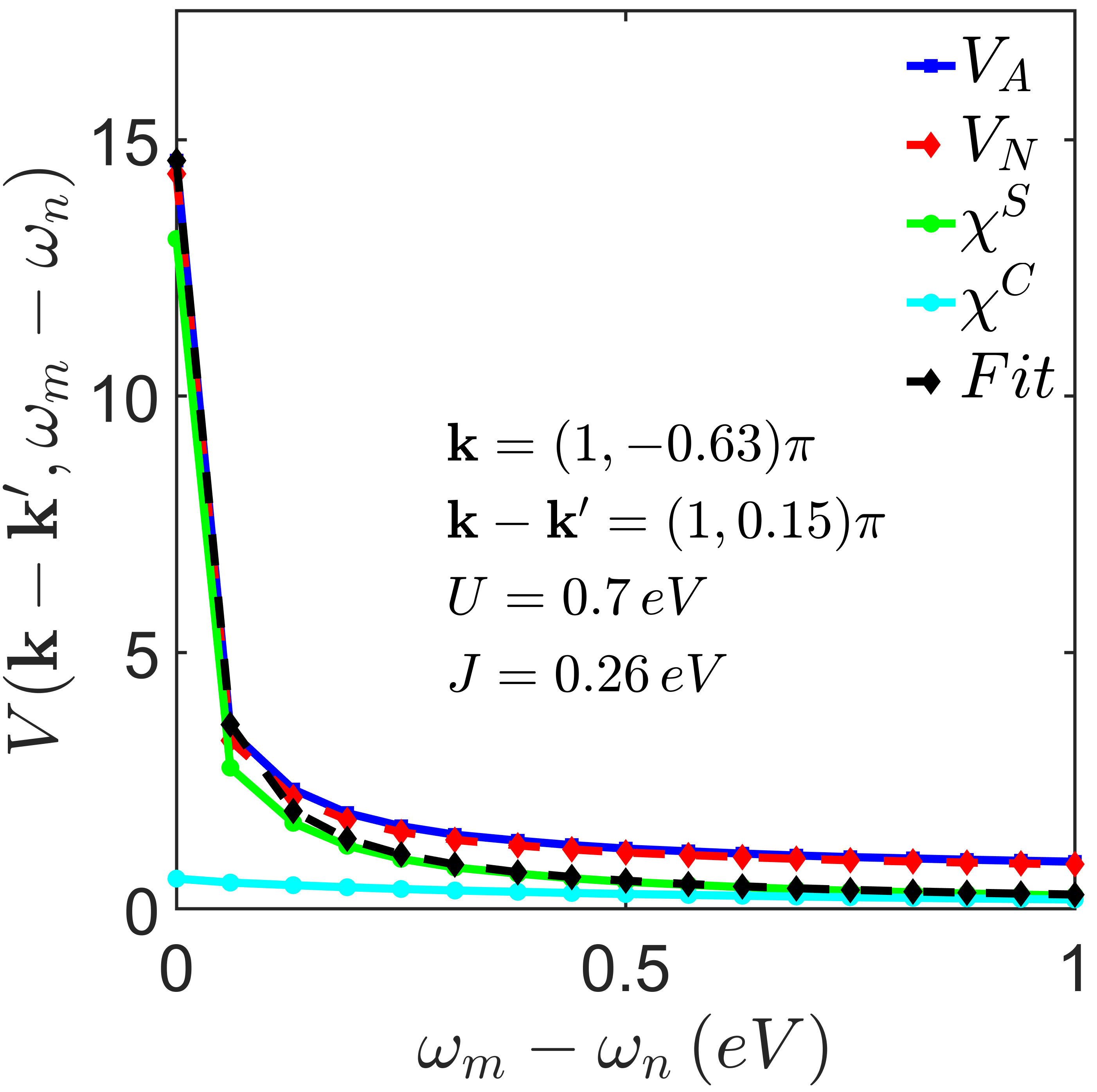} 
    \caption{The frequency dependence of the effective interaction $ V(\mathbf{k-k'},\omega_m-\omega_n)$}
    \label{fittedV}
\end{figure}
We will simplify the equations further, under assumptions at $T=T_C$, to obtain solutions without implementing numerically expensive self-consistency loops. The interaction vertices $ V_{\alpha\beta N/A}(\mathbf{k-k'},\omega_m-\omega_n)$ is expressed in terms of the spin fluctuation contribution $ \chi_{\alpha\beta}^S(\mathbf{k-k'},\omega_m-\omega_n)$ and the charge fluctuation contribution  $ \chi_{\alpha\beta}^C(\mathbf{k-k'},\omega_m-\omega_n)$. Let's denote $(\omega_m-\omega_n)$ by $\Omega_{mn}$. Fig.\ref{fittedV} shows the Matsubara frequency dependence of these two terms and of the interaction vertices for a momentum transfer of $(\pi,0.15\pi)$ (incommensurate antiferromagnetic wave-vector) connecting the electron and hole pocket at a typical interaction strength for our multiorbital pnictide system. Here, the spin fluctuation contribution is dominant and falls off on a frequency scale that is small compared with the bandwidth. Thus, the gap equation is dominated by important $\mathbf{k}$ and $\mathbf{k'}$ values restricted by this frequency cutoff to remain
near Fermi Surface. Without band crossing in the vicinity of the Fermi level, each FS point corresponds to a unique band quantum number, i.e. $\mathbf{k} \in \mu$, and $ \mathbf{k'} \in \nu$. Henceforth, we transfer the band index as a subscript of momentum index $\mathbf{k}$.

For $ \Omega_{mn} = 0 $ in the Matsubara axis, the real quantity $ V_{A/N}(\mathbf{k_\mu-k'_\nu},\Omega_{mn} = 0)$ translates to the real part of the Retarded quantity $ \text{Re}\left( V_{A/N}(\mathbf{k_\mu-k'_\nu},\Omega=0) \right). $ We assume the form of the interaction to have the following explicit dependence on the Matsubara frequency:
\begin{equation}
\label{V_Landaudamping}
V_{A/N}(\mathbf{k_\mu-k'_\nu},\Omega_{mn}) = \frac{  V_{A/N}(\mathbf{k_\mu-k'_\nu},\Omega_{mn}=0) }{ 1 + W |\Omega_{mn}| } = \frac{ \text{Re}\left( V_{A/N}(\mathbf{k_\mu-k'_\nu},\Omega=0) \right) }{ 1 + W| \omega_m - \omega_n | }
\end{equation}
where $W \propto \xi_C^{-1}$, $\xi_C$ being the spin-fluctuation energy cut-off. We show the band representation of the effective interaction vertices from orbital space in Eq.\ref{normalvertex} in the main text. In Fig.\ref{fittedV}, we also show the numerically fitted curve (in black) to the form of Eq.\ref{V_Landaudamping}. It shows good fit for low frequency region $ |\omega_m - \omega_n| < \xi_C $ which makes the largest contribution to the frequency sum in Eq.\ref{Eliash1} and \ref{Eliash2}. As for the external frequency, we solve the Eliashberg equations for low energy physics at the Fermi level at $ \omega_m = \omega_0=\pi k_B T_C$ (lowest Fermionic Matsubara frequency). This makes $ | \omega_m - \omega_n | = | \omega_0 - \omega_n | = | 2n\pi k_B T_C |$. For simplicity, we assume that the gap function is frequency independent $\Phi_\mu(\mathbf{k},\omega_n) = \Phi(\mathbf{k_\mu})$. 

First, we apply the above assumptions in $d$-dimensions to Eq.\ref{Eliash1}:
\begin{equation}
\label{intermediateZ}
    \omega_0 (1-Z(\mathbf{k_\mu})) = - \frac{1}{\beta (2\pi)^d} {\oint}_{\mathbf{k'_\nu} \in FS} \frac{\left[V(\mathbf{k_\mu,k'_\nu},0)\right]_N}{|\mathbf{v}_F(\mathbf{k'_\nu})|} (d^{d-1}k'_\nu)_\parallel \sum_{n=-\infty}^\infty \frac{\omega_n}{1+W|2n\pi k_B T_C|} \displaystyle{\int^{\infty}_{-\infty}} d\varepsilon'_{k'_\nu} \frac{1}{\omega_n^2 +  {\varepsilon'_{k'_\nu}}^2} 
\end{equation}
$\left[V(\mathbf{k_\mu,k'_\nu},0)\right]_N$ is expressed in Eq.\ref{normalvertex}. The momentum summation over full BZ has been reduced to a transverse and longitudinal integration over the Fermi surface values of the quantities involved in the equation. At $T_C$, we have $ \varepsilon(\mathbf{k'_\nu})  >> \Phi(\mathbf{k'_\nu}) $, hence $E(\mathbf{k'_\nu},\omega_n ) \approx \frac{ \varepsilon(\mathbf{k'_\nu})}{Z(\mathbf{k'_\nu})} = \varepsilon'_{k'_\nu} $. Since the effective interaction is peaked mostly for scattering vectors connecting states $(\mathbf{k_\mu,k'_\nu})$ on the FS, one can identify that the full BZ summation can be replaced by integration over small shells wrapped around the FS. Within the infinitesimal volume of each cubes surrounding individual FS points, we assume that there is only longitudinal variation of $ \left[V(\mathbf{k_\mu,k'_\nu},0)\right]_N$, $ Z(\mathbf{k'_\nu}) $ and $ \Phi(\mathbf{k'_\nu}) $ parallel to the FS, and no variation in the transverse direction. Hence, they are constant within this infinitesimal cube at any FS point $ k'_\nu$. $\varepsilon'_{k'_\nu}$ is constant along the FS and only varies in the transverse direction. The integration yields $\frac{2}{|\omega_n|} \text{tan}^{-1}\left(\infty\right) = \frac{\pi}{|\omega_n|}$ and together with the factor $1/\beta$ in the RHS it cancels the factor $\omega_0 = \pi k_B T_C = \pi/\beta $ in the LHS, while $\omega_n/|\omega_n|$ yields a $\text{sign}\left(\omega_n\right)$ in the RHS. Eq.\ref{intermediateZ} is rewritten as:
\begin{equation}
    Z(\mathbf{k_\mu}) = 1 + \frac{1}{ (2\pi)^d} \displaystyle{\oint_{\mathbf{k'_\nu} \in FS } } \left[V(\mathbf{k_\mu,k'_\nu},0)\right]_N \frac{(d^{d-1}k'_\nu)_\parallel}{|\mathbf{v}_F(\mathbf{k'_\nu})|} \sum_{n=-\infty}^\infty \frac{\text{sign}\left(\omega_n\right)}{1+W|2n\pi k_B T_C|} 
\end{equation}
The Matsubara sum just survives for $n=0$ and is cancelled out for all positive $(+n)$ and negative $(-n)$ pairs due to the oddness of $\text{sign}\left(\omega_n\right)$. Thus, with $\text{sign}\left(\omega_0\right) = +1$, we arrive at the following equation that needs to be numerically solved for $Z(\mathbf{k_\mu})$:
\begin{equation}
\label{BCS_Zeqn}
    \boxed{ Z(\mathbf{k_\mu}) = 1 + \frac{1}{ (2\pi)^d} \displaystyle{\oint_{\mathbf{k'_\nu} \in FS } } \left[V(\mathbf{k_\mu,k'_\nu},0)\right]_N \frac{(d^{d-1}k'_\nu)_\parallel}{|\mathbf{v}_F(\mathbf{k'_\nu})|} }
\end{equation}
Returning to Eq.\ref{Eliash2}, we proceed similarly as above:
\begin{equation}
\label{Linearized_eliash}
    \Phi(\mathbf{k_\mu}) = - \frac{\pi}{\beta (2\pi)^d} {\oint}_{\mathbf{k'_\nu} \in FS} \frac{\left[V(\mathbf{k_\mu,k'_\nu},0)\right]_A \Phi(\mathbf{k'_\nu})}{Z(\mathbf{k'_\nu})} \frac{(d^{d-1}k'_\nu)_\parallel}{|\mathbf{v}_F(\mathbf{k'_\nu})|} \sum_{n=-\infty}^\infty \frac{1}{1+W|2n\pi k_B T_C|}. \frac{1}{|\omega_n|} 
\end{equation}
We can perform the Matsubara sum. Rewriting in terms of the dimensionless parameter $ A=1/(2\pi W k_B T_C)$ :
\begin{equation}
    \frac{\pi}{\beta} \sum_{n=-\infty}^\infty \frac{1}{1+|n|/A}. \frac{1}{|\omega_n|} = \sum_{n=-\infty}^\infty \frac{A}{A+|n|}. \frac{1}{|2n+1|} = \sum_{n=0}^{\infty} \frac{A}{A+1+n} .\frac{1}{2n+1} + \sum_{n=0}^\infty \frac{A}{A+n}. \frac{1}{2n+1} 
\end{equation}
where we shifted $n$ to $(n-1)$ in the first infinite summation term. For low critical temperatures, $A>>1$, and we can absorb $A+1\approx A$ which gives:
\begin{equation}
    \begin{split}
    \label{Matsubara_sum}
    \frac{\pi}{\beta} \sum_{n=-\infty}^\infty \frac{1}{1+|n|/A}. \frac{1}{|\omega_n|} &= 2 \sum_{n=0}^\infty \frac{A}{A+n}. \frac{1}{2n+1} \\
    \end{split}
\end{equation}
We refer to identities of the Digamma function $ \Psi(z)$, one of which is:
\begin{equation}
    \begin{split}
    \Psi(z) &= - \gamma + \sum_{n=0}^\infty \frac{z-1}{(2n+1)(z+n)} \\
    \end{split}
\end{equation}
$\gamma$ is the Euler-Macheroni constant. For $z>>1$, we have $z-1 \approx z$ and $\Psi(z) \approx \text{log}(z)$. Thus, rewriting Eq.\ref{Matsubara_sum}:
\begin{equation}
    \begin{split}
    \frac{\pi}{\beta} \sum_{n=-\infty}^\infty \frac{1}{1+|n|/A}. \frac{1}{|\omega_n|} & \approx 2 \left(\gamma + \text{log}(A)\right) = 2 \text{log}\left(e^\gamma A\right)\\
    \end{split}
\end{equation}
This makes the linearized gap equation from Eq.\ref{Linearized_eliash} take the following form:
\begin{equation}
\label{unsym_gapeqn}
     \Phi(\mathbf{k_\mu}) = - \frac{ 2 \text{log}\left(e^\gamma A\right) }{(2\pi)^d} {\oint}_{\mathbf{k'_\nu} \in FS} \frac{\left[V(\mathbf{k_\mu,k'_\nu},0)\right]_A }{Z(\mathbf{k'_\nu})} \frac{(d^{d-1}k'_\nu)_\parallel}{|\mathbf{v}_F(\mathbf{k'_\nu})|} \Phi(\mathbf{k'_\nu}) 
\end{equation}
The gap equation above turns out to be an eigen value equation which can now be solved numerically, yielding eigen value $\lambda$ and vector $g(\mathbf{k})$. Multiplying with ${\oint}_{\mathbf{k_\mu} \in FS} \frac{\Phi(\mathbf{k_\mu})(d^{d-1}k_\mu)_\parallel}{(2\pi)^d|\mathbf{v}_F(\mathbf{k_\mu})|} $ on both sides of the above equation, we get:
\begin{equation}
\begin{split}
\label{eigenvalue_eqn}
     {\oint}_{\mathbf{k_\mu} \in FS} \frac{(d^{d-1}k_\mu)_\parallel}{(2\pi)^d|\mathbf{v}_F(\mathbf{k_\mu})|} \Phi^2(\mathbf{k_\mu}) &= - \frac{ 2 \text{log}\left(e^\gamma A\right) }{(2\pi)^{2d}} {\oint}_{\mathbf{k'_\nu} \in FS} {\oint}_{\mathbf{k_\mu} \in FS} \frac{(d^{d-1}k_\mu)_\parallel}{|\mathbf{v}_F(\mathbf{k_\mu})|} \Phi(\mathbf{k_\mu}) \frac{\left[V(\mathbf{k_\mu,k'_\nu},0)\right]_A }{Z(\mathbf{k'_\nu})} \frac{(d^{d-1}k'_\nu)_\parallel}{|\mathbf{v}_F(\mathbf{k'_\nu})|} \Phi(\mathbf{k'_\nu}) \\
\end{split}
\end{equation}
Defining $\lambda$ as:
\begin{equation}
     \lambda = - \frac{ \frac{ 1 }{(2\pi)^{2d}} {\oint}_{\mathbf{k'_\nu} \in FS} {\oint}_{\mathbf{k_\mu} \in FS} \frac{(d^{d-1}k_\mu)_\parallel}{|\mathbf{v}_F(\mathbf{k_\mu})|} \Phi(\mathbf{k_\mu}) \frac{\left[V(\mathbf{k_\mu,k'_\nu},0)\right]_A }{Z(\mathbf{k'_\nu})} \frac{(d^{d-1}k'_\nu)_\parallel}{|\mathbf{v}_F(\mathbf{k'_\nu})|} \Phi(\mathbf{k'_\nu}) }{{\oint}_{\mathbf{k_\mu} \in FS} \frac{(d^{d-1}k_\mu)_\parallel}{(2\pi)^d|\mathbf{v}_F(\mathbf{k_\mu})|} \Phi^2(\mathbf{k_\mu})}
     \label{lambda_equation}
\end{equation}
and dropping the constant log term from our numerical solution, we get the following eigen value equation from Eq.\ref{unsym_gapeqn} that needs to be numerically evaluated:

\begin{equation} \label{BCSgap}
 \boxed{ \lambda g(\mathbf{k_\mu}) 
 = -\frac{1}{ (2\pi)^d} \displaystyle{\sum_{\mathbf{k'_\nu} \in FS } } \frac{ 1 }{Z(\mathbf{k'_\nu}) } \left[V(\mathbf{k_\mu,k'_\nu},0)\right]_A \frac{(d^{d-1}k'_\nu)_\parallel}{|\mathbf{v}_F(\mathbf{k'_\nu})|}  g(\mathbf{k'_\nu}) }
\end{equation}
The final modified gap structure due to $ Z(\mathbf{k_\mu})$ is given by $ \Delta(\mathbf{k_\mu}) = \frac{g(\mathbf{k_\mu})}{Z(\mathbf{k_\mu})}$.

\bibliographystyle{apsrev4-2}
\bibliography{references}{}

\begin{thebibliography}{33}%
\makeatletter
\providecommand \@ifxundefined [1]{%
 \@ifx{#1\undefined}
}%
\providecommand \@ifnum [1]{%
 \ifnum #1\expandafter \@firstoftwo
 \else \expandafter \@secondoftwo
 \fi
}%
\providecommand \@ifx [1]{%
 \ifx #1\expandafter \@firstoftwo
 \else \expandafter \@secondoftwo
 \fi
}%
\providecommand \natexlab [1]{#1}%
\providecommand \enquote  [1]{``#1''}%
\providecommand \bibnamefont  [1]{#1}%
\providecommand \bibfnamefont [1]{#1}%
\providecommand \citenamefont [1]{#1}%
\providecommand \href@noop [0]{\@secondoftwo}%
\providecommand \href [0]{\begingroup \@sanitize@url \@href}%
\providecommand \@href[1]{\@@startlink{#1}\@@href}%
\providecommand \@@href[1]{\endgroup#1\@@endlink}%
\providecommand \@sanitize@url [0]{\catcode `\\12\catcode `\$12\catcode
  `\&12\catcode `\#12\catcode `\^12\catcode `\_12\catcode `\%12\relax}%
\providecommand \@@startlink[1]{}%
\providecommand \@@endlink[0]{}%
\providecommand \url  [0]{\begingroup\@sanitize@url \@url }%
\providecommand \@url [1]{\endgroup\@href {#1}{\urlprefix }}%
\providecommand \urlprefix  [0]{URL }%
\providecommand \Eprint [0]{\href }%
\providecommand \doibase [0]{https://doi.org/}%
\providecommand \selectlanguage [0]{\@gobble}%
\providecommand \bibinfo  [0]{\@secondoftwo}%
\providecommand \bibfield  [0]{\@secondoftwo}%
\providecommand \translation [1]{[#1]}%
\providecommand \BibitemOpen [0]{}%
\providecommand \bibitemStop [0]{}%
\providecommand \bibitemNoStop [0]{.\EOS\space}%
\providecommand \EOS [0]{\spacefactor3000\relax}%
\providecommand \BibitemShut  [1]{\csname bibitem#1\endcsname}%
\let\auto@bib@innerbib\@empty
\bibitem [{\citenamefont {de~Medici}(2015)}]{deMedici_review}%
  \BibitemOpen
  \bibfield  {author} {\bibinfo {author} {\bibfnamefont {L.}~\bibnamefont
  {de~Medici}},\ }\href {https://doi.org/10.1007/978-3-319-11254-1} {\emph
  {\bibinfo {title} {Iron-Based Superconductivity, Weak and Strong Correlations
  in {Fe} Superconductors}}},\ edited by\ \bibinfo {editor} {\bibnamefont
  {{Peter D. Johnson, Guangyong Xu, and Wei-Guo Yin}}},\ Springer Series in
  Materials Science\ (\bibinfo  {publisher} {Springer},\ \bibinfo {year}
  {2015})\BibitemShut {NoStop}%
\bibitem [{\citenamefont {Bascones}\ \emph {et~al.}(2016)\citenamefont
  {Bascones}, \citenamefont {Valenzuela},\ and\ \citenamefont
  {Calder\'on}}]{Bascones_review}%
  \BibitemOpen
  \bibfield  {author} {\bibinfo {author} {\bibfnamefont {E.}~\bibnamefont
  {Bascones}}, \bibinfo {author} {\bibfnamefont {B.}~\bibnamefont
  {Valenzuela}},\ and\ \bibinfo {author} {\bibfnamefont {M.~J.}\ \bibnamefont
  {Calder\'on}},\ }\href
  {https://doi.org/http://dx.doi.org/10.1016/j.crhy.2015.05.004} {\bibfield
  {journal} {\bibinfo  {journal} {C. R. Phys.}\ }\textbf {\bibinfo {volume}
  {17}},\ \bibinfo {pages} {36} (\bibinfo {year} {2016})},\ \bibinfo {note}
  {iron-based superconductors / Supraconducteurs \`a base de fer}\BibitemShut
  {NoStop}%
\bibitem [{\citenamefont {van Roekeghem}\ \emph {et~al.}(2016)\citenamefont
  {van Roekeghem}, \citenamefont {Richard}, \citenamefont {Ding},\ and\
  \citenamefont {Biermann}}]{Biermann_review}%
  \BibitemOpen
  \bibfield  {author} {\bibinfo {author} {\bibfnamefont {A.}~\bibnamefont {van
  Roekeghem}}, \bibinfo {author} {\bibfnamefont {P.}~\bibnamefont {Richard}},
  \bibinfo {author} {\bibfnamefont {H.}~\bibnamefont {Ding}},\ and\ \bibinfo
  {author} {\bibfnamefont {S.}~\bibnamefont {Biermann}},\ }\href
  {https://doi.org/http://dx.doi.org/10.1016/j.crhy.2015.11.003} {\bibfield
  {journal} {\bibinfo  {journal} {C. R. Phys.}\ }\textbf {\bibinfo {volume}
  {17}},\ \bibinfo {pages} {140} (\bibinfo {year} {2016})},\ \bibinfo {note}
  {iron-based superconductors / Supraconducteurs \`a base de fer}\BibitemShut
  {NoStop}%
\bibitem [{\citenamefont {Yin}\ \emph {et~al.}(2011)\citenamefont {Yin},
  \citenamefont {Haule},\ and\ \citenamefont {Kotliar}}]{Yin2011}%
  \BibitemOpen
  \bibfield  {author} {\bibinfo {author} {\bibfnamefont {Z.~P.}\ \bibnamefont
  {Yin}}, \bibinfo {author} {\bibfnamefont {K.}~\bibnamefont {Haule}},\ and\
  \bibinfo {author} {\bibfnamefont {G.}~\bibnamefont {Kotliar}},\ }\href
  {https://doi.org/10.1038/nmat3120} {\bibfield  {journal} {\bibinfo  {journal}
  {Nat. Mater.}\ }\textbf {\bibinfo {volume} {10}},\ \bibinfo {pages} {932}
  (\bibinfo {year} {2011})}\BibitemShut {NoStop}%
\bibitem [{\citenamefont {Li}\ \emph {et~al.}(2016)\citenamefont {Li},
  \citenamefont {Yin}, \citenamefont {Wang}, \citenamefont {Tam}, \citenamefont
  {Abernathy}, \citenamefont {Podlesnyak}, \citenamefont {Zhang}, \citenamefont
  {Wang}, \citenamefont {Xing}, \citenamefont {Jin}, \citenamefont {Haule},
  \citenamefont {Kotliar}, \citenamefont {Maier},\ and\ \citenamefont
  {Dai}}]{Li_orb_sel_16}%
  \BibitemOpen
  \bibfield  {author} {\bibinfo {author} {\bibfnamefont {Y.}~\bibnamefont
  {Li}}, \bibinfo {author} {\bibfnamefont {Z.}~\bibnamefont {Yin}}, \bibinfo
  {author} {\bibfnamefont {X.}~\bibnamefont {Wang}}, \bibinfo {author}
  {\bibfnamefont {D.~W.}\ \bibnamefont {Tam}}, \bibinfo {author} {\bibfnamefont
  {D.~L.}\ \bibnamefont {Abernathy}}, \bibinfo {author} {\bibfnamefont
  {A.}~\bibnamefont {Podlesnyak}}, \bibinfo {author} {\bibfnamefont
  {C.}~\bibnamefont {Zhang}}, \bibinfo {author} {\bibfnamefont
  {M.}~\bibnamefont {Wang}}, \bibinfo {author} {\bibfnamefont {L.}~\bibnamefont
  {Xing}}, \bibinfo {author} {\bibfnamefont {C.}~\bibnamefont {Jin}}, \bibinfo
  {author} {\bibfnamefont {K.}~\bibnamefont {Haule}}, \bibinfo {author}
  {\bibfnamefont {G.}~\bibnamefont {Kotliar}}, \bibinfo {author} {\bibfnamefont
  {T.~A.}\ \bibnamefont {Maier}},\ and\ \bibinfo {author} {\bibfnamefont
  {P.}~\bibnamefont {Dai}},\ }\href
  {https://doi.org/10.1103/PhysRevLett.116.247001} {\bibfield  {journal}
  {\bibinfo  {journal} {Phys. Rev. Lett.}\ }\textbf {\bibinfo {volume} {116}},\
  \bibinfo {pages} {247001} (\bibinfo {year} {2016})}\BibitemShut {NoStop}%
\bibitem [{\citenamefont {Ye}\ \emph {et~al.}(2014)\citenamefont {Ye},
  \citenamefont {Zhang}, \citenamefont {Chen}, \citenamefont {Xu},
  \citenamefont {Jiang}, \citenamefont {Niu}, \citenamefont {Wen},
  \citenamefont {Xing}, \citenamefont {Wang}, \citenamefont {Jin},
  \citenamefont {Xie},\ and\ \citenamefont {Feng}}]{Ye_doping_FeSc_14}%
  \BibitemOpen
  \bibfield  {author} {\bibinfo {author} {\bibfnamefont {Z.~R.}\ \bibnamefont
  {Ye}}, \bibinfo {author} {\bibfnamefont {Y.}~\bibnamefont {Zhang}}, \bibinfo
  {author} {\bibfnamefont {F.}~\bibnamefont {Chen}}, \bibinfo {author}
  {\bibfnamefont {M.}~\bibnamefont {Xu}}, \bibinfo {author} {\bibfnamefont
  {J.}~\bibnamefont {Jiang}}, \bibinfo {author} {\bibfnamefont {X.~H.}\
  \bibnamefont {Niu}}, \bibinfo {author} {\bibfnamefont {C.~H.~P.}\
  \bibnamefont {Wen}}, \bibinfo {author} {\bibfnamefont {L.~Y.}\ \bibnamefont
  {Xing}}, \bibinfo {author} {\bibfnamefont {X.~C.}\ \bibnamefont {Wang}},
  \bibinfo {author} {\bibfnamefont {C.~Q.}\ \bibnamefont {Jin}}, \bibinfo
  {author} {\bibfnamefont {B.~P.}\ \bibnamefont {Xie}},\ and\ \bibinfo {author}
  {\bibfnamefont {D.~L.}\ \bibnamefont {Feng}},\ }\href
  {https://doi.org/10.1103/PhysRevX.4.031041} {\bibfield  {journal} {\bibinfo
  {journal} {Phys. Rev. X}\ }\textbf {\bibinfo {volume} {4}},\ \bibinfo {pages}
  {031041} (\bibinfo {year} {2014})}\BibitemShut {NoStop}%
\bibitem [{\citenamefont {Arakawa}\ and\ \citenamefont
  {Ogata}(2011)}]{Ogata_selectivepairing}%
  \BibitemOpen
  \bibfield  {author} {\bibinfo {author} {\bibfnamefont {N.}~\bibnamefont
  {Arakawa}}\ and\ \bibinfo {author} {\bibfnamefont {M.}~\bibnamefont
  {Ogata}},\ }\href {https://doi.org/10.1143/JPSJ.80.074704} {\bibfield
  {journal} {\bibinfo  {journal} {J. Phys. Soc. Jpn.}\ }\textbf {\bibinfo
  {volume} {80}},\ \bibinfo {pages} {074704} (\bibinfo {year}
  {2011})}\BibitemShut {NoStop}%
\bibitem [{\citenamefont {Yu}\ \emph {et~al.}(2014)\citenamefont {Yu},
  \citenamefont {Zhu},\ and\ \citenamefont {Si}}]{Si_selectivepairing}%
  \BibitemOpen
  \bibfield  {author} {\bibinfo {author} {\bibfnamefont {R.}~\bibnamefont
  {Yu}}, \bibinfo {author} {\bibfnamefont {J.-X.}\ \bibnamefont {Zhu}},\ and\
  \bibinfo {author} {\bibfnamefont {Q.}~\bibnamefont {Si}},\ }\href
  {https://doi.org/10.1103/PhysRevB.89.024509} {\bibfield  {journal} {\bibinfo
  {journal} {Phys. Rev. B}\ }\textbf {\bibinfo {volume} {89}},\ \bibinfo
  {pages} {024509} (\bibinfo {year} {2014})}\BibitemShut {NoStop}%
\bibitem [{\citenamefont {Yin}\ \emph {et~al.}(2014)\citenamefont {Yin},
  \citenamefont {Haule},\ and\ \citenamefont {Kotliar}}]{Yin2014}%
  \BibitemOpen
  \bibfield  {author} {\bibinfo {author} {\bibfnamefont {Z.~P.}\ \bibnamefont
  {Yin}}, \bibinfo {author} {\bibfnamefont {K.}~\bibnamefont {Haule}},\ and\
  \bibinfo {author} {\bibfnamefont {G.}~\bibnamefont {Kotliar}},\ }\href
  {http://dx.doi.org/10.1038/nphys3116} {\bibfield  {journal} {\bibinfo
  {journal} {Nat. Phys.}\ }\textbf {\bibinfo {volume} {10}},\ \bibinfo {pages}
  {845} (\bibinfo {year} {2014})}\BibitemShut {NoStop}%
\bibitem [{\citenamefont {Sprau}\ \emph {et~al.}(2017)\citenamefont {Sprau},
  \citenamefont {Kostin}, \citenamefont {Kreisel}, \citenamefont {B{\"o}hmer},
  \citenamefont {Taufour}, \citenamefont {Canfield}, \citenamefont {Mukherjee},
  \citenamefont {Hirschfeld}, \citenamefont {Andersen},\ and\ \citenamefont
  {Davis}}]{Sprau2017}%
  \BibitemOpen
  \bibfield  {author} {\bibinfo {author} {\bibfnamefont {P.~O.}\ \bibnamefont
  {Sprau}}, \bibinfo {author} {\bibfnamefont {A.}~\bibnamefont {Kostin}},
  \bibinfo {author} {\bibfnamefont {A.}~\bibnamefont {Kreisel}}, \bibinfo
  {author} {\bibfnamefont {A.~E.}\ \bibnamefont {B{\"o}hmer}}, \bibinfo
  {author} {\bibfnamefont {V.}~\bibnamefont {Taufour}}, \bibinfo {author}
  {\bibfnamefont {P.~C.}\ \bibnamefont {Canfield}}, \bibinfo {author}
  {\bibfnamefont {S.}~\bibnamefont {Mukherjee}}, \bibinfo {author}
  {\bibfnamefont {P.~J.}\ \bibnamefont {Hirschfeld}}, \bibinfo {author}
  {\bibfnamefont {B.~M.}\ \bibnamefont {Andersen}},\ and\ \bibinfo {author}
  {\bibfnamefont {J.~C.~S.}\ \bibnamefont {Davis}},\ }\href
  {https://doi.org/10.1126/science.aal1575} {\bibfield  {journal} {\bibinfo
  {journal} {Science}\ }\textbf {\bibinfo {volume} {357}},\ \bibinfo {pages}
  {75} (\bibinfo {year} {2017})},\ \Eprint
  {https://arxiv.org/abs/http://science.sciencemag.org/content/357/6346/75.full.pdf}
  {http://science.sciencemag.org/content/357/6346/75.full.pdf} \BibitemShut
  {NoStop}%
\bibitem [{\citenamefont {Scalapino}(2012)}]{Scalapino2012}%
  \BibitemOpen
  \bibfield  {author} {\bibinfo {author} {\bibfnamefont {D.~J.}\ \bibnamefont
  {Scalapino}},\ }\href {https://doi.org/10.1103/RevModPhys.84.1383} {\bibfield
   {journal} {\bibinfo  {journal} {Rev. Mod. Phys.}\ }\textbf {\bibinfo
  {volume} {84}},\ \bibinfo {pages} {1383} (\bibinfo {year}
  {2012})}\BibitemShut {NoStop}%
\bibitem [{\citenamefont {Hirschfeld}(2016)}]{HirschfeldCRAS}%
  \BibitemOpen
  \bibfield  {author} {\bibinfo {author} {\bibfnamefont {P.~J.}\ \bibnamefont
  {Hirschfeld}},\ }\href
  {https://doi.org/http://dx.doi.org/10.1016/j.crhy.2015.10.002} {\bibfield
  {journal} {\bibinfo  {journal} {C. R. Phys.}\ }\textbf {\bibinfo {volume}
  {17}},\ \bibinfo {pages} {197} (\bibinfo {year} {2016})},\ \bibinfo {note}
  {iron-based superconductors / Supraconducteurs à base de fer}\BibitemShut
  {NoStop}%
\bibitem [{\citenamefont {Chubukov}(2015)}]{Chubukov_review}%
  \BibitemOpen
  \bibfield  {author} {\bibinfo {author} {\bibfnamefont {A.}~\bibnamefont
  {Chubukov}},\ }\href {https://doi.org/10.1007/978-3-319-11254-1} {\emph
  {\bibinfo {title} {Iron-Based Superconductivity, Itinerant electron scenario
  for Fe-based superconductors}}},\ edited by\ \bibinfo {editor} {\bibnamefont
  {{Peter D. Johnson, Guangyong Xu, and Wei-Guo Yin}}},\ Springer Series in
  Materials Science\ (\bibinfo  {publisher} {Springer},\ \bibinfo {year}
  {2015})\BibitemShut {NoStop}%
\bibitem [{\citenamefont {Mazin}\ \emph {et~al.}(2008)\citenamefont {Mazin},
  \citenamefont {Singh}, \citenamefont {Johannes},\ and\ \citenamefont
  {Du}}]{Mazin2008}%
  \BibitemOpen
  \bibfield  {author} {\bibinfo {author} {\bibfnamefont {I.~I.}\ \bibnamefont
  {Mazin}}, \bibinfo {author} {\bibfnamefont {D.~J.}\ \bibnamefont {Singh}},
  \bibinfo {author} {\bibfnamefont {M.~D.}\ \bibnamefont {Johannes}},\ and\
  \bibinfo {author} {\bibfnamefont {M.~H.}\ \bibnamefont {Du}},\ }\href
  {https://doi.org/10.1103/PhysRevLett.101.057003} {\bibfield  {journal}
  {\bibinfo  {journal} {Phys. Rev. Lett.}\ }\textbf {\bibinfo {volume} {101}},\
  \bibinfo {pages} {057003} (\bibinfo {year} {2008})}\BibitemShut {NoStop}%
\bibitem [{\citenamefont {Maier}\ \emph {et~al.}(2009)\citenamefont {Maier},
  \citenamefont {Graser}, \citenamefont {Scalapino},\ and\ \citenamefont
  {Hirschfeld}}]{Maier_anisotropy_2009}%
  \BibitemOpen
  \bibfield  {author} {\bibinfo {author} {\bibfnamefont {T.~A.}\ \bibnamefont
  {Maier}}, \bibinfo {author} {\bibfnamefont {S.}~\bibnamefont {Graser}},
  \bibinfo {author} {\bibfnamefont {D.~J.}\ \bibnamefont {Scalapino}},\ and\
  \bibinfo {author} {\bibfnamefont {P.~J.}\ \bibnamefont {Hirschfeld}},\ }\href
  {https://doi.org/10.1103/PhysRevB.79.224510} {\bibfield  {journal} {\bibinfo
  {journal} {Phys. Rev. B}\ }\textbf {\bibinfo {volume} {79}},\ \bibinfo
  {pages} {224510} (\bibinfo {year} {2009})}\BibitemShut {NoStop}%
\bibitem [{\citenamefont {Kreisel}\ \emph {et~al.}(2017)\citenamefont
  {Kreisel}, \citenamefont {Andersen}, \citenamefont {Sprau}, \citenamefont
  {Kostin}, \citenamefont {Davis},\ and\ \citenamefont
  {Hirschfeld}}]{Kreisel2017}%
  \BibitemOpen
  \bibfield  {author} {\bibinfo {author} {\bibfnamefont {A.}~\bibnamefont
  {Kreisel}}, \bibinfo {author} {\bibfnamefont {B.~M.}\ \bibnamefont
  {Andersen}}, \bibinfo {author} {\bibfnamefont {P.~O.}\ \bibnamefont {Sprau}},
  \bibinfo {author} {\bibfnamefont {A.}~\bibnamefont {Kostin}}, \bibinfo
  {author} {\bibfnamefont {J.~C.~S.}\ \bibnamefont {Davis}},\ and\ \bibinfo
  {author} {\bibfnamefont {P.~J.}\ \bibnamefont {Hirschfeld}},\ }\href
  {https://doi.org/10.1103/PhysRevB.95.174504} {\bibfield  {journal} {\bibinfo
  {journal} {Phys. Rev. B}\ }\textbf {\bibinfo {volume} {95}},\ \bibinfo
  {pages} {174504} (\bibinfo {year} {2017})}\BibitemShut {NoStop}%
\bibitem [{\citenamefont {Lee}\ \emph {et~al.}(2012)\citenamefont {Lee},
  \citenamefont {Ji}, \citenamefont {Kim}, \citenamefont {Kim}, \citenamefont
  {Haule}, \citenamefont {Kotliar}, \citenamefont {Lee}, \citenamefont {Khim},
  \citenamefont {Kim}, \citenamefont {Kim}, \citenamefont {Kim},\ and\
  \citenamefont {Shim}}]{Lee_etal_Kotliar2012}%
  \BibitemOpen
  \bibfield  {author} {\bibinfo {author} {\bibfnamefont {G.}~\bibnamefont
  {Lee}}, \bibinfo {author} {\bibfnamefont {H.~S.}\ \bibnamefont {Ji}},
  \bibinfo {author} {\bibfnamefont {Y.}~\bibnamefont {Kim}}, \bibinfo {author}
  {\bibfnamefont {C.}~\bibnamefont {Kim}}, \bibinfo {author} {\bibfnamefont
  {K.}~\bibnamefont {Haule}}, \bibinfo {author} {\bibfnamefont
  {G.}~\bibnamefont {Kotliar}}, \bibinfo {author} {\bibfnamefont
  {B.}~\bibnamefont {Lee}}, \bibinfo {author} {\bibfnamefont {S.}~\bibnamefont
  {Khim}}, \bibinfo {author} {\bibfnamefont {K.~H.}\ \bibnamefont {Kim}},
  \bibinfo {author} {\bibfnamefont {K.~S.}\ \bibnamefont {Kim}}, \bibinfo
  {author} {\bibfnamefont {K.-S.}\ \bibnamefont {Kim}},\ and\ \bibinfo {author}
  {\bibfnamefont {J.~H.}\ \bibnamefont {Shim}},\ }\href
  {https://doi.org/10.1103/PhysRevLett.109.177001} {\bibfield  {journal}
  {\bibinfo  {journal} {Phys. Rev. Lett.}\ }\textbf {\bibinfo {volume} {109}},\
  \bibinfo {pages} {177001} (\bibinfo {year} {2012})}\BibitemShut {NoStop}%
\bibitem [{\citenamefont {Ferber}\ \emph {et~al.}(2012)\citenamefont {Ferber},
  \citenamefont {Foyevtsova}, \citenamefont {Valent\'{\i}},\ and\ \citenamefont
  {Jeschke}}]{Ferber2012}%
  \BibitemOpen
  \bibfield  {author} {\bibinfo {author} {\bibfnamefont {J.}~\bibnamefont
  {Ferber}}, \bibinfo {author} {\bibfnamefont {K.}~\bibnamefont {Foyevtsova}},
  \bibinfo {author} {\bibfnamefont {R.}~\bibnamefont {Valent\'{\i}}},\ and\
  \bibinfo {author} {\bibfnamefont {H.~O.}\ \bibnamefont {Jeschke}},\ }\href
  {https://doi.org/10.1103/PhysRevB.85.094505} {\bibfield  {journal} {\bibinfo
  {journal} {Phys. Rev. B}\ }\textbf {\bibinfo {volume} {85}},\ \bibinfo
  {pages} {094505} (\bibinfo {year} {2012})}\BibitemShut {NoStop}%
\bibitem [{\citenamefont {Borisenko}\ \emph {et~al.}(2010)\citenamefont
  {Borisenko}, \citenamefont {Zabolotnyy}, \citenamefont {Evtushinsky},
  \citenamefont {Kim}, \citenamefont {Morozov}, \citenamefont {Yaresko},
  \citenamefont {Kordyuk}, \citenamefont {Behr}, \citenamefont {Vasiliev},
  \citenamefont {Follath},\ and\ \citenamefont {B\"uchner}}]{BorisenkoLiFeAs}%
  \BibitemOpen
  \bibfield  {author} {\bibinfo {author} {\bibfnamefont {S.~V.}\ \bibnamefont
  {Borisenko}}, \bibinfo {author} {\bibfnamefont {V.~B.}\ \bibnamefont
  {Zabolotnyy}}, \bibinfo {author} {\bibfnamefont {D.~V.}\ \bibnamefont
  {Evtushinsky}}, \bibinfo {author} {\bibfnamefont {T.~K.}\ \bibnamefont
  {Kim}}, \bibinfo {author} {\bibfnamefont {I.~V.}\ \bibnamefont {Morozov}},
  \bibinfo {author} {\bibfnamefont {A.~N.}\ \bibnamefont {Yaresko}}, \bibinfo
  {author} {\bibfnamefont {A.~A.}\ \bibnamefont {Kordyuk}}, \bibinfo {author}
  {\bibfnamefont {G.}~\bibnamefont {Behr}}, \bibinfo {author} {\bibfnamefont
  {A.}~\bibnamefont {Vasiliev}}, \bibinfo {author} {\bibfnamefont
  {R.}~\bibnamefont {Follath}},\ and\ \bibinfo {author} {\bibfnamefont
  {B.}~\bibnamefont {B\"uchner}},\ }\href
  {https://doi.org/10.1103/PhysRevLett.105.067002} {\bibfield  {journal}
  {\bibinfo  {journal} {Phys. Rev. Lett.}\ }\textbf {\bibinfo {volume} {105}},\
  \bibinfo {pages} {067002} (\bibinfo {year} {2010})}\BibitemShut {NoStop}%
\bibitem [{\citenamefont {Borisenko}\ \emph {et~al.}(2012)\citenamefont
  {Borisenko}, \citenamefont {Zabolotnyy}, \citenamefont {Kordyuk},
  \citenamefont {Evtushinsky}, \citenamefont {Kim}, \citenamefont {Morozov},
  \citenamefont {Follath},\ and\ \citenamefont {Büchner}}]{Borisenko12}%
  \BibitemOpen
  \bibfield  {author} {\bibinfo {author} {\bibfnamefont {S.~V.}\ \bibnamefont
  {Borisenko}}, \bibinfo {author} {\bibfnamefont {V.~B.}\ \bibnamefont
  {Zabolotnyy}}, \bibinfo {author} {\bibfnamefont {A.~A.}\ \bibnamefont
  {Kordyuk}}, \bibinfo {author} {\bibfnamefont {D.~V.}\ \bibnamefont
  {Evtushinsky}}, \bibinfo {author} {\bibfnamefont {T.~K.}\ \bibnamefont
  {Kim}}, \bibinfo {author} {\bibfnamefont {I.~V.}\ \bibnamefont {Morozov}},
  \bibinfo {author} {\bibfnamefont {R.}~\bibnamefont {Follath}},\ and\ \bibinfo
  {author} {\bibfnamefont {B.}~\bibnamefont {Büchner}},\ }\href
  {https://doi.org/10.3390/sym4010251} {\bibfield  {journal} {\bibinfo
  {journal} {Symmetry}\ }\textbf {\bibinfo {volume} {4}},\ \bibinfo {pages}
  {251} (\bibinfo {year} {2012})}\BibitemShut {NoStop}%
\bibitem [{\citenamefont {Zantout}\ \emph {et~al.}(2019)\citenamefont
  {Zantout}, \citenamefont {Backes},\ and\ \citenamefont
  {Valent\'{\i}}}]{ZantoutPRL2019}%
  \BibitemOpen
  \bibfield  {author} {\bibinfo {author} {\bibfnamefont {K.}~\bibnamefont
  {Zantout}}, \bibinfo {author} {\bibfnamefont {S.}~\bibnamefont {Backes}},\
  and\ \bibinfo {author} {\bibfnamefont {R.}~\bibnamefont {Valent\'{\i}}},\
  }\href {https://doi.org/10.1103/PhysRevLett.123.256401} {\bibfield  {journal}
  {\bibinfo  {journal} {Phys. Rev. Lett.}\ }\textbf {\bibinfo {volume} {123}},\
  \bibinfo {pages} {256401} (\bibinfo {year} {2019})}\BibitemShut {NoStop}%
\bibitem [{\citenamefont {Platt}\ \emph {et~al.}(2011)\citenamefont {Platt},
  \citenamefont {Thomale},\ and\ \citenamefont {Hanke}}]{Thomale2011}%
  \BibitemOpen
  \bibfield  {author} {\bibinfo {author} {\bibfnamefont {C.}~\bibnamefont
  {Platt}}, \bibinfo {author} {\bibfnamefont {R.}~\bibnamefont {Thomale}},\
  and\ \bibinfo {author} {\bibfnamefont {W.}~\bibnamefont {Hanke}},\ }\href
  {https://doi.org/10.1103/PhysRevB.84.235121} {\bibfield  {journal} {\bibinfo
  {journal} {Phys. Rev. B}\ }\textbf {\bibinfo {volume} {84}},\ \bibinfo
  {pages} {235121} (\bibinfo {year} {2011})}\BibitemShut {NoStop}%
\bibitem [{\citenamefont {Wang}\ \emph {et~al.}(2013)\citenamefont {Wang},
  \citenamefont {Kreisel}, \citenamefont {Zabolotnyy}, \citenamefont
  {Borisenko}, \citenamefont {B\"uchner}, \citenamefont {Maier}, \citenamefont
  {Hirschfeld},\ and\ \citenamefont {Scalapino}}]{Wang13}%
  \BibitemOpen
  \bibfield  {author} {\bibinfo {author} {\bibfnamefont {Y.}~\bibnamefont
  {Wang}}, \bibinfo {author} {\bibfnamefont {A.}~\bibnamefont {Kreisel}},
  \bibinfo {author} {\bibfnamefont {V.~B.}\ \bibnamefont {Zabolotnyy}},
  \bibinfo {author} {\bibfnamefont {S.~V.}\ \bibnamefont {Borisenko}}, \bibinfo
  {author} {\bibfnamefont {B.}~\bibnamefont {B\"uchner}}, \bibinfo {author}
  {\bibfnamefont {T.~A.}\ \bibnamefont {Maier}}, \bibinfo {author}
  {\bibfnamefont {P.~J.}\ \bibnamefont {Hirschfeld}},\ and\ \bibinfo {author}
  {\bibfnamefont {D.~J.}\ \bibnamefont {Scalapino}},\ }\href
  {https://doi.org/10.1103/PhysRevB.88.174516} {\bibfield  {journal} {\bibinfo
  {journal} {Phys. Rev. B}\ }\textbf {\bibinfo {volume} {88}},\ \bibinfo
  {pages} {174516} (\bibinfo {year} {2013})}\BibitemShut {NoStop}%
\bibitem [{\citenamefont {Saito}\ \emph {et~al.}(2014)\citenamefont {Saito},
  \citenamefont {Onari}, \citenamefont {Yamakawa}, \citenamefont {Kontani},
  \citenamefont {Borisenko},\ and\ \citenamefont {Zabolotnyy}}]{Saito14}%
  \BibitemOpen
  \bibfield  {author} {\bibinfo {author} {\bibfnamefont {T.}~\bibnamefont
  {Saito}}, \bibinfo {author} {\bibfnamefont {S.}~\bibnamefont {Onari}},
  \bibinfo {author} {\bibfnamefont {Y.}~\bibnamefont {Yamakawa}}, \bibinfo
  {author} {\bibfnamefont {H.}~\bibnamefont {Kontani}}, \bibinfo {author}
  {\bibfnamefont {S.~V.}\ \bibnamefont {Borisenko}},\ and\ \bibinfo {author}
  {\bibfnamefont {V.~B.}\ \bibnamefont {Zabolotnyy}},\ }\href
  {https://doi.org/10.1103/PhysRevB.90.035104} {\bibfield  {journal} {\bibinfo
  {journal} {Phys. Rev. B}\ }\textbf {\bibinfo {volume} {90}},\ \bibinfo
  {pages} {035104} (\bibinfo {year} {2014})}\BibitemShut {NoStop}%
\bibitem [{\citenamefont {Ahn}\ \emph {et~al.}(2014)\citenamefont {Ahn},
  \citenamefont {Eremin}, \citenamefont {Knolle}, \citenamefont {Zabolotnyy},
  \citenamefont {Borisenko}, \citenamefont {B\"uchner},\ and\ \citenamefont
  {Chubukov}}]{Ahn14}%
  \BibitemOpen
  \bibfield  {author} {\bibinfo {author} {\bibfnamefont {F.}~\bibnamefont
  {Ahn}}, \bibinfo {author} {\bibfnamefont {I.}~\bibnamefont {Eremin}},
  \bibinfo {author} {\bibfnamefont {J.}~\bibnamefont {Knolle}}, \bibinfo
  {author} {\bibfnamefont {V.~B.}\ \bibnamefont {Zabolotnyy}}, \bibinfo
  {author} {\bibfnamefont {S.~V.}\ \bibnamefont {Borisenko}}, \bibinfo {author}
  {\bibfnamefont {B.}~\bibnamefont {B\"uchner}},\ and\ \bibinfo {author}
  {\bibfnamefont {A.~V.}\ \bibnamefont {Chubukov}},\ }\href
  {https://doi.org/10.1103/PhysRevB.89.144513} {\bibfield  {journal} {\bibinfo
  {journal} {Phys. Rev. B}\ }\textbf {\bibinfo {volume} {89}},\ \bibinfo
  {pages} {144513} (\bibinfo {year} {2014})}\BibitemShut {NoStop}%
\bibitem [{\citenamefont {Allan}\ \emph {et~al.}(2012)\citenamefont {Allan},
  \citenamefont {Rost}, \citenamefont {Mackenzie}, \citenamefont {Xie},
  \citenamefont {Davis}, \citenamefont {Kihou}, \citenamefont {Lee},
  \citenamefont {Iyo}, \citenamefont {Eisaki},\ and\ \citenamefont
  {Chuang}}]{Allan12}%
  \BibitemOpen
  \bibfield  {author} {\bibinfo {author} {\bibfnamefont {M.~P.}\ \bibnamefont
  {Allan}}, \bibinfo {author} {\bibfnamefont {A.~W.}\ \bibnamefont {Rost}},
  \bibinfo {author} {\bibfnamefont {A.~P.}\ \bibnamefont {Mackenzie}}, \bibinfo
  {author} {\bibfnamefont {Y.}~\bibnamefont {Xie}}, \bibinfo {author}
  {\bibfnamefont {J.~C.}\ \bibnamefont {Davis}}, \bibinfo {author}
  {\bibfnamefont {K.}~\bibnamefont {Kihou}}, \bibinfo {author} {\bibfnamefont
  {C.~H.}\ \bibnamefont {Lee}}, \bibinfo {author} {\bibfnamefont
  {A.}~\bibnamefont {Iyo}}, \bibinfo {author} {\bibfnamefont {H.}~\bibnamefont
  {Eisaki}},\ and\ \bibinfo {author} {\bibfnamefont {T.-M.}\ \bibnamefont
  {Chuang}},\ }\href {https://doi.org/10.1126/science.1218726} {\bibfield
  {journal} {\bibinfo  {journal} {Science}\ }\textbf {\bibinfo {volume}
  {336}},\ \bibinfo {pages} {563} (\bibinfo {year} {2012})}\BibitemShut
  {NoStop}%
\bibitem [{\citenamefont {Umezawa}\ \emph {et~al.}(2012)\citenamefont
  {Umezawa}, \citenamefont {Li}, \citenamefont {Miao}, \citenamefont
  {Nakayama}, \citenamefont {Liu}, \citenamefont {Richard}, \citenamefont
  {Sato}, \citenamefont {He}, \citenamefont {Wang}, \citenamefont {Chen},
  \citenamefont {Ding}, \citenamefont {Takahashi},\ and\ \citenamefont
  {Wang}}]{Umezawa12}%
  \BibitemOpen
  \bibfield  {author} {\bibinfo {author} {\bibfnamefont {K.}~\bibnamefont
  {Umezawa}}, \bibinfo {author} {\bibfnamefont {Y.}~\bibnamefont {Li}},
  \bibinfo {author} {\bibfnamefont {H.}~\bibnamefont {Miao}}, \bibinfo {author}
  {\bibfnamefont {K.}~\bibnamefont {Nakayama}}, \bibinfo {author}
  {\bibfnamefont {Z.-H.}\ \bibnamefont {Liu}}, \bibinfo {author} {\bibfnamefont
  {P.}~\bibnamefont {Richard}}, \bibinfo {author} {\bibfnamefont
  {T.}~\bibnamefont {Sato}}, \bibinfo {author} {\bibfnamefont {J.~B.}\
  \bibnamefont {He}}, \bibinfo {author} {\bibfnamefont {D.-M.}\ \bibnamefont
  {Wang}}, \bibinfo {author} {\bibfnamefont {G.~F.}\ \bibnamefont {Chen}},
  \bibinfo {author} {\bibfnamefont {H.}~\bibnamefont {Ding}}, \bibinfo {author}
  {\bibfnamefont {T.}~\bibnamefont {Takahashi}},\ and\ \bibinfo {author}
  {\bibfnamefont {S.-C.}\ \bibnamefont {Wang}},\ }\href
  {https://doi.org/10.1103/PhysRevLett.108.037002} {\bibfield  {journal}
  {\bibinfo  {journal} {Phys. Rev. Lett.}\ }\textbf {\bibinfo {volume} {108}},\
  \bibinfo {pages} {037002} (\bibinfo {year} {2012})}\BibitemShut {NoStop}%
\bibitem [{\citenamefont {Ikeda}(2008)}]{Ikeda2008}%
  \BibitemOpen
  \bibfield  {author} {\bibinfo {author} {\bibfnamefont {H.}~\bibnamefont
  {Ikeda}},\ }\href {https://doi.org/10.1143/JPSJ.77.123707} {\bibfield
  {journal} {\bibinfo  {journal} {Journal of the Physical Society of Japan}\
  }\textbf {\bibinfo {volume} {77}},\ \bibinfo {pages} {123707} (\bibinfo
  {year} {2008})},\ \Eprint
  {https://arxiv.org/abs/https://doi.org/10.1143/JPSJ.77.123707}
  {https://doi.org/10.1143/JPSJ.77.123707} \BibitemShut {NoStop}%
\bibitem [{\citenamefont {Ikeda}\ \emph {et~al.}(2010)\citenamefont {Ikeda},
  \citenamefont {Arita},\ and\ \citenamefont {Kune\ifmmode~\check{s}\else
  \v{s}\fi{}}}]{AritaIkeda2010}%
  \BibitemOpen
  \bibfield  {author} {\bibinfo {author} {\bibfnamefont {H.}~\bibnamefont
  {Ikeda}}, \bibinfo {author} {\bibfnamefont {R.}~\bibnamefont {Arita}},\ and\
  \bibinfo {author} {\bibfnamefont {J.}~\bibnamefont
  {Kune\ifmmode~\check{s}\else \v{s}\fi{}}},\ }\href
  {https://doi.org/10.1103/PhysRevB.81.054502} {\bibfield  {journal} {\bibinfo
  {journal} {Phys. Rev. B}\ }\textbf {\bibinfo {volume} {81}},\ \bibinfo
  {pages} {054502} (\bibinfo {year} {2010})}\BibitemShut {NoStop}%
\bibitem [{\citenamefont {Yanagi}\ \emph {et~al.}(2010)\citenamefont {Yanagi},
  \citenamefont {Yamakawa}, \citenamefont {Adachi},\ and\ \citenamefont
  {\ifmmode~\bar{O}\else \={O}\fi{}no}}]{Yanagi2010}%
  \BibitemOpen
  \bibfield  {author} {\bibinfo {author} {\bibfnamefont {Y.}~\bibnamefont
  {Yanagi}}, \bibinfo {author} {\bibfnamefont {Y.}~\bibnamefont {Yamakawa}},
  \bibinfo {author} {\bibfnamefont {N.}~\bibnamefont {Adachi}},\ and\ \bibinfo
  {author} {\bibfnamefont {Y.}~\bibnamefont {\ifmmode~\bar{O}\else
  \={O}\fi{}no}},\ }\href {https://doi.org/10.1103/PhysRevB.82.064518}
  {\bibfield  {journal} {\bibinfo  {journal} {Phys. Rev. B}\ }\textbf {\bibinfo
  {volume} {82}},\ \bibinfo {pages} {064518} (\bibinfo {year}
  {2010})}\BibitemShut {NoStop}%
\bibitem [{\citenamefont {Kuroki}\ \emph {et~al.}(2008)\citenamefont {Kuroki},
  \citenamefont {Onari}, \citenamefont {Arita}, \citenamefont {Usui},
  \citenamefont {Tanaka}, \citenamefont {Kontani},\ and\ \citenamefont
  {Aoki}}]{Kuroki2008}%
  \BibitemOpen
  \bibfield  {author} {\bibinfo {author} {\bibfnamefont {K.}~\bibnamefont
  {Kuroki}}, \bibinfo {author} {\bibfnamefont {S.}~\bibnamefont {Onari}},
  \bibinfo {author} {\bibfnamefont {R.}~\bibnamefont {Arita}}, \bibinfo
  {author} {\bibfnamefont {H.}~\bibnamefont {Usui}}, \bibinfo {author}
  {\bibfnamefont {Y.}~\bibnamefont {Tanaka}}, \bibinfo {author} {\bibfnamefont
  {H.}~\bibnamefont {Kontani}},\ and\ \bibinfo {author} {\bibfnamefont
  {H.}~\bibnamefont {Aoki}},\ }\href
  {https://doi.org/10.1103/PhysRevLett.101.087004} {\bibfield  {journal}
  {\bibinfo  {journal} {Phys. Rev. Lett.}\ }\textbf {\bibinfo {volume} {101}},\
  \bibinfo {pages} {087004} (\bibinfo {year} {2008})}\BibitemShut {NoStop}%
\bibitem [{\citenamefont {Triola}\ \emph {et~al.}(2020)\citenamefont {Triola},
  \citenamefont {Cayao},\ and\ \citenamefont {Black-Schaffer}}]{Schaffer2020}%
  \BibitemOpen
  \bibfield  {author} {\bibinfo {author} {\bibfnamefont {C.}~\bibnamefont
  {Triola}}, \bibinfo {author} {\bibfnamefont {J.}~\bibnamefont {Cayao}},\ and\
  \bibinfo {author} {\bibfnamefont {A.~M.}\ \bibnamefont {Black-Schaffer}},\
  }\href {https://doi.org/10.1002/andp.201900298} {\bibfield  {journal}
  {\bibinfo  {journal} {Annalen der Physik}\ }\textbf {\bibinfo {volume}
  {532}},\ \bibinfo {pages} {1900298} (\bibinfo {year} {2020})},\ \Eprint
  {https://arxiv.org/abs/https://onlinelibrary.wiley.com/doi/pdf/10.1002/andp.201900298}
  {https://onlinelibrary.wiley.com/doi/pdf/10.1002/andp.201900298} \BibitemShut
  {NoStop}%
\bibitem [{\citenamefont {Maiti}\ \emph {et~al.}(2011)\citenamefont {Maiti},
  \citenamefont {Korshunov}, \citenamefont {Maier}, \citenamefont
  {Hirschfeld},\ and\ \citenamefont {Chubukov}}]{Maiti2011}%
  \BibitemOpen
  \bibfield  {author} {\bibinfo {author} {\bibfnamefont {S.}~\bibnamefont
  {Maiti}}, \bibinfo {author} {\bibfnamefont {M.~M.}\ \bibnamefont
  {Korshunov}}, \bibinfo {author} {\bibfnamefont {T.~A.}\ \bibnamefont
  {Maier}}, \bibinfo {author} {\bibfnamefont {P.~J.}\ \bibnamefont
  {Hirschfeld}},\ and\ \bibinfo {author} {\bibfnamefont {A.~V.}\ \bibnamefont
  {Chubukov}},\ }\href {https://doi.org/10.1103/PhysRevLett.107.147002}
  {\bibfield  {journal} {\bibinfo  {journal} {Phys. Rev. Lett.}\ }\textbf
  {\bibinfo {volume} {107}},\ \bibinfo {pages} {147002} (\bibinfo {year}
  {2011})}\BibitemShut {NoStop}%
\end{thebibliography}%

\end{document}